\documentclass[trackchanges]{aastex701}

\usepackage[T1]{fontenc}

\DeclareRobustCommand{\VAN}[3]{#2}
\let\VANthebibliography\thebibliography
\def\thebibliography{\DeclareRobustCommand{\VAN}[3]{##3}\VANthebibliography}

\usepackage{mathtools}  

\usepackage{multirow}
\usepackage{natbib}
\usepackage{subcaption}
\usepackage{tabularx}  

\PassOptionsToPackage{table,xcdraw}{xcolor}
\usepackage{graphicx}
\usepackage{tikz}
\usepackage{xcolor}
\usepackage{colortbl}

\usepackage{siunitx} 
\usepackage{booktabs}

\definecolor{myBlue}{HTML}{1f77b4}    
\definecolor{myRed}{HTML}{e74c3c}     
\definecolor{myGreen}{HTML}{2ca02c}   
\definecolor{myMaroon}{HTML}{d62728}  
\definecolor{myPurple}{HTML}{9b59b6}  
\definecolor{myOrange}{HTML}{f39c12}  
\usepackage[toc]{appendix}

\usepackage{hyperref} 

\begin{document}

\title{Attention-Based Preprocessing Framework for Improving Rare Transient Classification}

\author[orcid=0000-0002-6527-1368,sname='Sheng']{Xinyue Sheng}
\altaffiliation{Astrophysics Research Centre, School of Mathematics and Physics}
\affiliation{Queen's University Belfast, Northern Ireland, United Kingdom}
\email[show]{x.sheng@qub.ac.uk}  

\author[orcid=0000-0002-2183-4640,gname=Dung, sname='Pham']{Tuan Dung Pham} 
\altaffiliation{School of Electronics, Electrical Engineering and Computer Science}
\affiliation{Queen's University Belfast, Northern Ireland, United Kingdom}
\email{tpham01@qub.ac.uk}

\author[0009-0008-2472-689X,gname=Zichi,sname=Zhang]{Zichi Zhang}
\altaffiliation{School of Electronics, Electrical Engineering and Computer Science}
\affiliation{Queen's University Belfast, Northern Ireland, United Kingdom}
\email{zzhang54@qub.ac.uk}

\author[0000-0002-2555-3192,gname=Matt,sname=Nicholl]{Matt Nicholl}
\altaffiliation{Astrophysics Research Centre, School of Mathematics and Physics}
\affiliation{Queen's University Belfast, Northern Ireland, United Kingdom}
\email{matt.nicholl@qub.ac.uk}

\author[0000-0003-4599-1525,gname=Son, sname=Mai]{Thai Son Mai}
\altaffiliation{School of Electronics, Electrical Engineering and Computer Science}
\affiliation{Queen's University Belfast, Northern Ireland, United Kingdom}
\email{thaison.mai@qub.ac.uk}

\begin{abstract}
With large numbers of transients discovered by current and future imaging surveys, machine learning is increasingly applied to light curve and host galaxy properties to select events for follow-up. However, finding rare types of transients remains difficult due to extreme class imbalances in training sets, and extracting features from host images is complicated by the presence of bright foreground sources, particularly if the true host is faint or distant. Here we present a data augmentation pipeline for images and light curves that mitigates these issues, and apply this to improve classification of Superluminous Supernovae Type I (SLSNe-I) and Tidal Disruption Events (TDEs) with our existing NEEDLE code. The method uses a Similarity Index to remove image artefacts, and a masking procedure that removes unrelated sources while preserving the transient and its host. This focuses classifier attention on the relevant pixels, and enables arbitrary rotations for class upsampling. We also fit observed multi-band light curves with a two-dimensional Gaussian Process and generate data-driven synthetic samples by resampling and redshifting these models, cross-matching with galaxy images in the same class to produce unique but realistic new examples for training. Models trained with the augmented dataset achieve substantially higher purity: for classifications with a confidence of 0.8 or higher, we achieve 75\% (43\%) purity and 75\% (66\%) completeness for SLSNe-I (TDEs). 
\end{abstract}



\section{Introduction}\label{sec:intro}

Machine learning (ML) algorithms have been increasingly and successfully applied to astronomical data for a wide range of tasks.
In the field of transient identification, however, several challenges persist -- most notably, extremely imbalanced class distributions, varying image quality, and irregular observational cadence. These issues are common in large-scale survey data and pose significant obstacles to effective modelling, particularly for rare event classification, as highlighted in several studies (e.g. \citealt{Gomez_2020, Gomez_2023, Aleo_2024}).
While class imbalance usually arises because of physical differences in the volumetric rates of different types of transients, a range of factors can affect the quality of astronomical images. These issues can arise due to astronomical effects, such as saturated stars; telescope/hardware limitations, such as chip gaps; data reduction, including \texttt{NaN} values or inconsistent image sizes; or combinations of these effects, e.g. diffraction spikes. 

We recently developed the NEural Engine for Discovering Luminous Events (\texttt{NEEDLE}; \citealt{Sheng_2024}), a hybrid model using images and light curves to classify rare Type I superluminous supernovae (SLSNe-I) and tidal disruption events (TDEs), trained on data from the Zwicky Transient Facility (ZTF) Bright Transient Survey \citep{Bellm_2019,Perley_2020}.
Of the ZTF images used in the \texttt{NEEDLE} training set, $\gtrsim1.7$\% were affected by image artefacts that prevented their use in training the model -- and would prevent future objects with the same artefacts from being classified successfully. Nevertheless, these images still contain valuable information for classification tasks and should not be disregarded. Even for images without artefacts, machine learning suffers from large numbers of unrelated foreground/background sources in any astronomical image, which have no physical association with the transient of interest.

Additionally, the issue of extreme class imbalance becomes particularly evident in codes like \texttt{NEEDLE} that search for rare events. The training set from \citet{Sheng_2024} contained 5237 normal Supernovae (SN Ia and core-collapse SN), but only 87 SLSNe-I and 64 TDEs.
To address class imbalance in training data, two main strategies have emerged in the literature. \emph{Model-Centric Approaches} focus on enhancing model architecture and training techniques. Recent advances include the use of Transformer-based architectures \citep{Moreno-Cartagena_2023, Donoso-Oliva_2023, Chen_2023, Bairouk_2023, Allam_2023, Allam_2024, Cabrera-Vives_2024} and the development of custom loss functions that place greater emphasis on under-represented classes \citep{Villar_2023, Shah_2025}. \emph{Data-Centric Approaches} on the other hand aim to improve the quality and diversity of the training data itself. Strategies include the use of simulated datasets such as \textit{PLAsTiCC} and \textit{ELAsTiCC} \citep{Narayan_2023}, or statistical models, e.g. Supernova
Parametric Model (SPM), to reproduce the shape of light curves \citep{Pimentel_2023}. Other methods focus on domain-specific feature extraction and data modelling prior to training \citep{Pasquet_2019, Boone_2021}, as well as data augmentation techniques like Synthetic Minority Oversampling Technique (SMOTE, \citealt{Chawla_2002}), used in applications such as \citet{Dauphin_2020}.

Many successful transient classification pipelines combine both model- and data-centric approaches. However, recent empirical studies \citep[e.g.,][]{Whang_2020, Bhatt_2024} indicate that improvements in data quality can lead to more significant performance gains than model enhancements alone. These findings underscore the critical role of data-centric strategies in advancing classification accuracy. 
However, augmenting data sets with simulated data requires caution: if these simulations are not representative of the full range of transients in nature, models trained on them may exhibit systematic biases and poor generalization. Simulations therefore demand a thorough understanding of the underlying population and access to high-fidelity survey data for comparison.
Similarly, light curve simulations derived from physical models can also fail to adequately represent the full range of rare events. 
For example, in the \textit{PLAsTiCC} dataset \citep{Kessler_2019}, some of the physical models used were informed by only a handful of real events.

Motivated by these insights, our work focuses on enhancing the diversity and representativeness of training data--independent of further model complexity. Our objective is to guide feature learning toward more informative representations by supplying higher-quality, survey-consistent input samples. 
To this end, we introduce a scalable data preprocessing framework for transient classification that integrates both imaging and light curve information. Unlike many existing approaches, our method generates augmented training data exclusively from real observations, without relying on simulations or physical models. This ensures that the augmented dataset remains faithful to the underlying characteristics of the survey, while improving the model's ability to generalize to rare and complex transient phenomena. As part of this approach, we also remove artefacts and unrelated nearby sources from transient images -- this improves our ability to upsample these images without reinforcing spurious features, while also helping the network to focus attention on the transient and its host galaxy.

The structure of the paper is as follows:
Section~\ref{section1} revisits the architecture of \texttt{NEEDLE1.0}, detailing its upgrades and performance on real-time ZTF alerts.
Section~\ref{section2} introduces our methods for image restoration, masking, and up-sampling.
Section~\ref{section3} describes the re-sampling strategies applied to light curve data.
Section~\ref{sec:crossmatching} presents a novel cross-matching approach that augments the training set by combining same-class image and light curve data.
Section~\ref{section5} reports the results of ablation studies evaluating the impact of different augmentation methods.
Finally, Section~\ref{section6} provides a broader discussion and summarizes our main conclusions.

\section{NEEDLE 1.0 upgrades and performance}\label{section1}

\subsection{NEEDLE1.0 wrap-up}
\texttt{NEEDLE1.0} is a multi-modal classifier that identifies SLSNe-I and TDEs at their early stages. The motivation is to boost rare transient discovery and spectroscopic follow-up, to get a full physical picture of individual events and build a sample for statistical analysis. \texttt{NEEDLE} combines Convolutional Neural Networks (CNN) for imaging inputs and Fully-Dense Neural Networks (DNN) for metadata inputs. The logic is that \texttt{NEEDLE} utilizes context information: host galaxy offset, host colours, and morphology, alongside transient light curves, offering a full feature pattern to help find SLSN-I and TDEs effectively. 
The current version of \texttt{NEEDLE} has been trained on the ZTF bright transient survey \citep{Perley_2020}. 
A full description of the \texttt{NEEDLE} algorithm, training and early performance is given by \citealt{Sheng_2024}.

\subsection{Updates: additional features and weightings}

In its initial version, \texttt{NEEDLE} utilized only the $r$-band transient light curves, as $r$-band observations outnumber those in the $g$ and $i$ bands. In October 2024, \texttt{NEEDLE} was upgraded to incorporate additional bands when available. This results in a larger number of light curve features for classification and training. Of course, these features are not always independent: an event with a long rise time in the $r$ band will likely also have a long rise in the $g$ band, for example.
The updated metadata is shown in Table~\ref{tab:updated-metadata}.

Normalization and standardization are essential steps in data pre-processing. These are defined following:
\begin{equation}
\begin{aligned}
\text{Standardization:} \quad & x' = \frac{x - \mu}{\sigma} \\
\text{Normalization:} \quad & x' = \frac{x - \min(x)}{\max(x) - \min(x)}
\end{aligned}
\label{eq:preprocessing}
\end{equation}
where $x$ is the value of the feature, $\mu$ is the population mean, and $\sigma$ the standard deviation. These techniques are commonly used to preserve the shape of the feature distribution while rescaling values into a fixed range, which facilitates efficient computation within neural networks. They also significantly influence the training performance.

\begin{figure*}
    \centering
    \includegraphics[width=\linewidth]{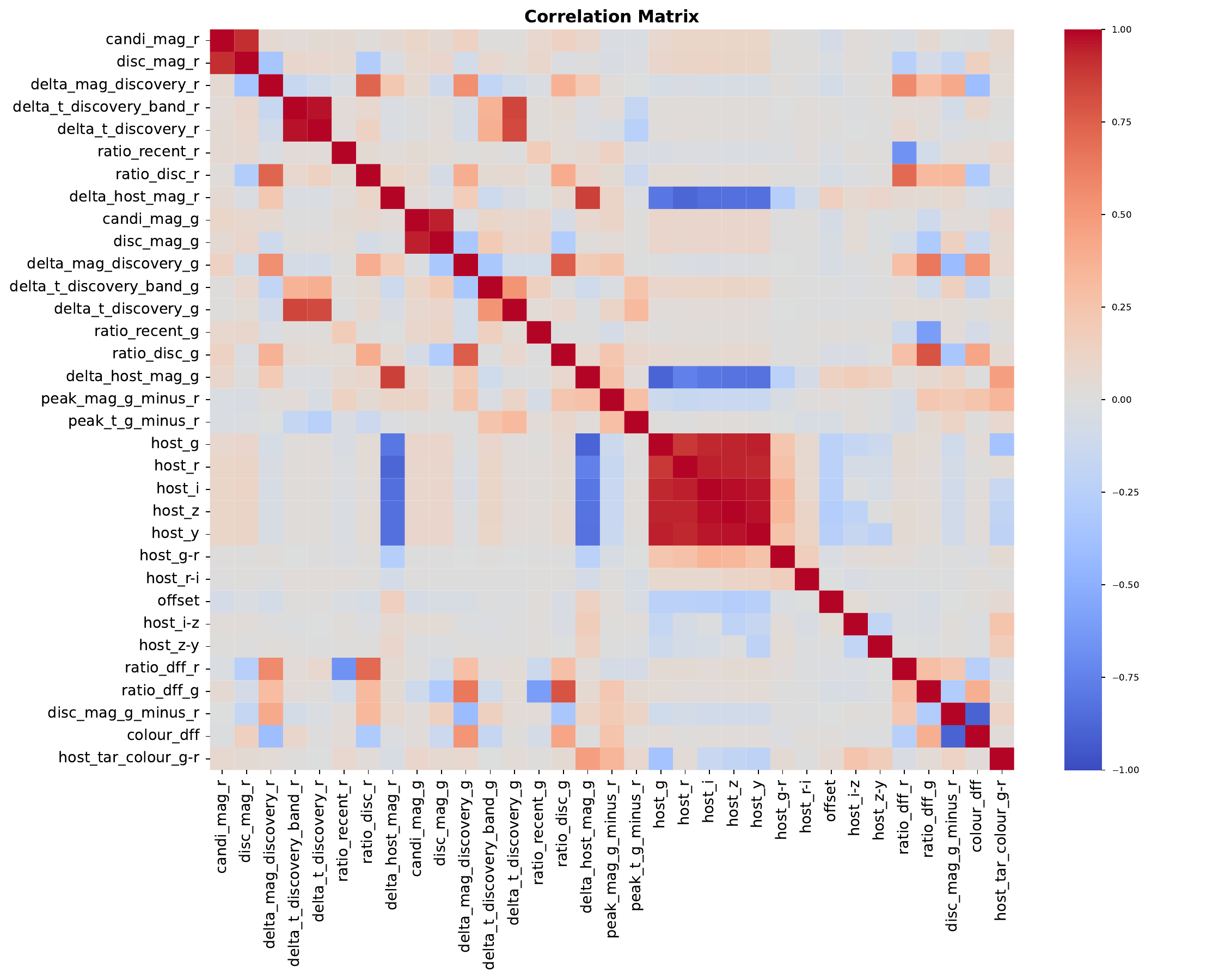}
    \caption{Feature correlations in NEEDLE after adding new features from subtractions of related features.}
    \label{fig:upgrate-correlation}
\end{figure*}

Input scaling can be performed in different ways: feature-wise or sample-wise. Feature-wise scaling may hinder comparison between features within a single sample, whereas sample-wise scaling can distort the distribution across features \citep{de_Amorim_2022}. 
To mitigate potential issues arising from inconsistent normalization strategies, we analysed pairwise feature correlations, as shown in Figure~\ref{fig:upgrate-correlation}.
For highly correlated feature pairs, we computed their differences and added these as new input features (shown with olive colour in Table \ref{tab:updated-metadata}), aiming to enhance the network's ability to capture subtle relational patterns across features.

Deployed via the PyPI package \texttt{xgboost}, \textit{eXtreme Gradient Boosting} (XGBoost; \citealt{Chen_2016}) is a scalable and efficient implementation of gradient-boosted decision trees. It constructs an ensemble of decision trees sequentially, with each new tree trained to correct the errors of its predecessors. Although \texttt{XGBoost} is highly effective for structured data, neural networks are more suitable for\texttt{NEEDLE}, which combines both tabular metadata and image inputs.

\begin{figure*}
    
    \centering
    \includegraphics[width=\linewidth]
    {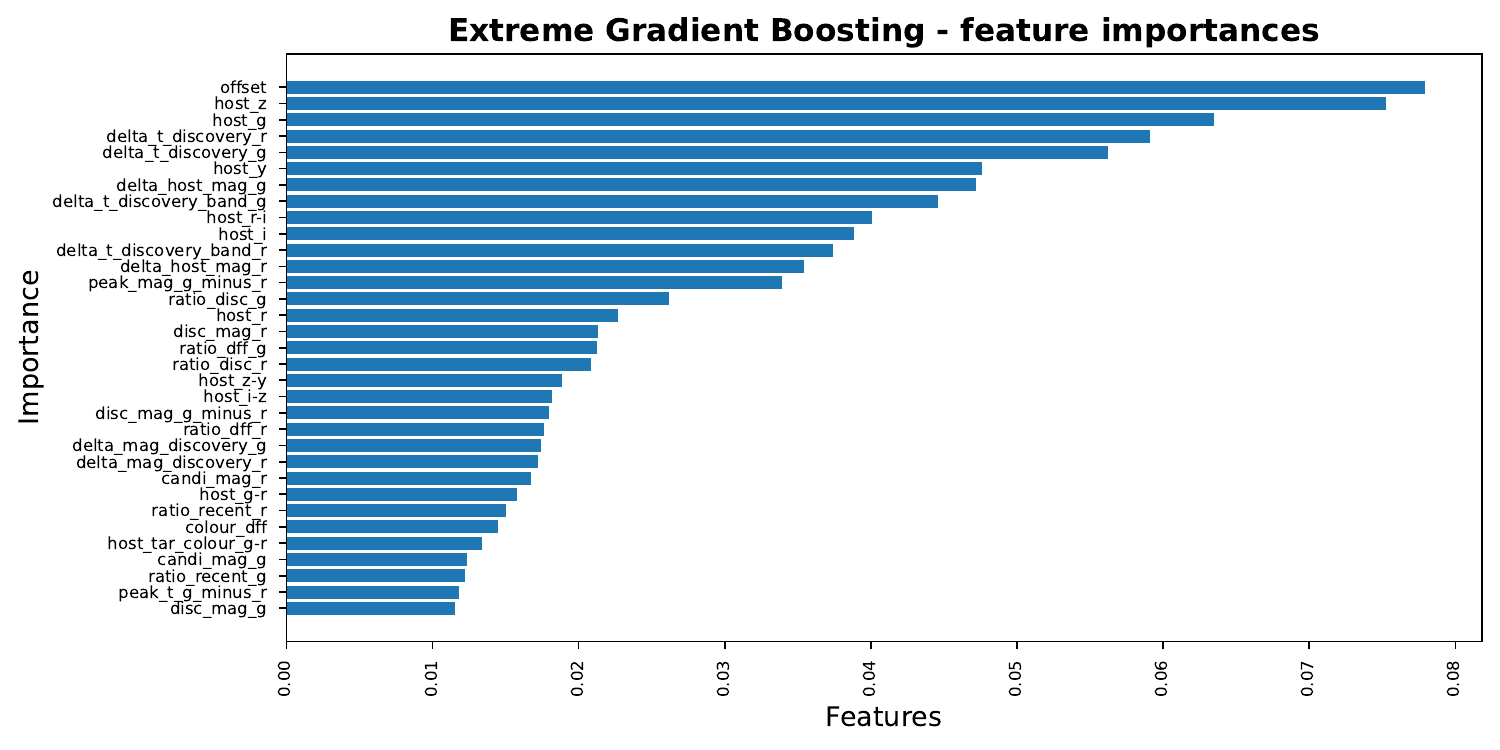}
    \caption{XGB feature ranking for NEEDLE metadata features.}
    \label{fig:xgb-ranking}
\end{figure*}

A key advantage of \texttt{XGBoost} is its ability to provide feature-importance rankings based on their contributions to classification. These rankings can be directly used to design a feature-weighting input layer ahead of the fully connected layers in a neural network architecture. As shown in Figure~\ref{fig:xgb-ranking}, the most important features include the offset and host magnitude in multiple bands, followed by rise time and the transient–host contrast. The colours of the transient and host appear next in importance. These features are critical for distinguishing SLSNe-I and TDEs: both exhibit bluer colours and longer rise times than normal SNe; TDEs preferentially occur near the centres of (often green) galaxies with small offsets; and SLSNe-I are typically found in faint, blue galaxies. These features are highly ranked, confirming their role as the key to the identification.

\begin{table}[]
\centering
\resizebox{\columnwidth}{!}{
\begin{tabular}{cc}
\multicolumn{2}{c}{\cellcolor[HTML]{CBCEFB}\textbf{Transient Features}} \\
\multicolumn{1}{c|}{\textbf{Feature Name}} &
  \textbf{Definition} \\
\multicolumn{1}{c|}{$m_{peak, r}$} &
  Peak magnitude in $r$ band among all existing \\
\multicolumn{1}{c|}{$m_{discovery, r}$} &
  Magnitude of the discovery observation in $r$ band. \\
\multicolumn{1}{c|}{$\Delta m_{discovery, r}$} &
  Magnitude difference between current observation and discovery observation in $r$ band. \\
\multicolumn{1}{c|}{$\Delta T_{discovery, r}$} &
  Time difference between current observation and earliest observation in $r$ band. \\
\multicolumn{1}{c|}{$\Delta T_{discovery, r*}$} &
  Time difference between current observation in $r$ band and discovery observation in $g$ or $r$ band. \\
\multicolumn{1}{c|}{$\lambda_{r,recent}=\Delta m_{recent, r} / \Delta T_{recent, r}$} &
  Ratio of magnitude difference over time difference since last detection in $r$ band \\
\multicolumn{1}{c|}{$\lambda_{r, discovery}=\Delta m_{discovery,r} / \Delta T_{discovery, r}$} &
  Ratio of magnitude difference over time difference since the earliest in $r$ band. \\
\multicolumn{1}{c|}{$\Delta m_{host-peak, r}$} &
  Magnitude difference between the $m_{host,r}$ and $m_{peak,r}$ \\
\multicolumn{1}{c|}{$m_{peak, g}$} &
  Peak magnitude in $g$ band among all existing \\
\multicolumn{1}{c|}{$m_{discovery, g}$} &
  Magnitude of the discovery observation in $g$ band. \\
\multicolumn{1}{c|}{$\Delta m_{discovery, g}$} &
  Magnitude difference between current observation and earliest observation in $g$ band. \\
\multicolumn{1}{c|}{$\Delta T_{discovery, g}$} &
  Time difference between current observation and earliest observation in $g$ band. \\
\multicolumn{1}{c|}{$\Delta T_{discovery, g*}$} &
  Time difference between current observation in $g$ band and discovery observation in $g$ or $r$ band. \\
\multicolumn{1}{c|}{$\lambda_{g, recent}=\Delta m_{recent, g} / \Delta T_{recent, g}$} &
  Ratio of magnitude difference over time difference since last detection in $g$ or $r$ band \\
\multicolumn{1}{c|}{$\lambda_{g, discovery}=\Delta m_{discovery,g} / \Delta T_{discovery, g}$} &
  Ratio of magnitude difference over time difference since the earliest in $g$ band \\
\multicolumn{1}{c|}{$\Delta m_{host-peak, g}$} &
  Magnitude difference between the $m_{host,g}$ and $m_{peak,g}$ \\
\multicolumn{1}{c|}{$\Delta m_{peak, g-r}$} &
  Magnitude difference between $g$ and $r$ bands around peak \\
\multicolumn{1}{c|}{$\Delta T_{peak, g-r}$} &
  Time difference between $g$ and $r$ bands around peak \\
\multicolumn{1}{c|}{\textit{separationArcsec (offset)}} &
  The distance (arcsec) between the host centre and transient on the image \\
\multicolumn{1}{c|}{\textcolor{olive}{$\lambda_{r, discovery} - \lambda_{r, recent}$}} &
  Ratio difference between $\lambda_{r, discovery}$ and $\lambda_{r, recent}$ \\
\multicolumn{1}{c|}{\textcolor{olive}{$\lambda_{g, discovery} - \lambda_{g, recent}$}}&
  Ratio difference between $\lambda_{g, discovery}$ and $\lambda_{g, recent}$ \\
\multicolumn{1}{c|}{\textcolor{olive}{$\Delta m_{discovery, g-r}$}} &
  Magnitude difference between the first $g$ and $r$ band detections \\
\multicolumn{1}{c|}{\textcolor{olive}{$\Delta \textit{colour}_{transient}$}} &
  Colour difference between the peak and discovery: $\Delta m_{peak, g-r}$ - $\Delta m_{discovery, g-r}$ \\
\multicolumn{2}{c}{\cellcolor[HTML]{CBCEFB}\textbf{Host Metadata Features}} \\
\multicolumn{1}{c|}{\textbf{Feature Name}} &
  \textbf{Definition} \\
\multicolumn{1}{c|}{$m_{gAp}$} &
  Host magnitude in $g$ band \\
\multicolumn{1}{c|}{$m_{rAp}$} &
  Host magnitude in $r$ band \\
\multicolumn{1}{c|}{$m_{iAp}$} &
  Host magnitude in $i$ band \\
\multicolumn{1}{c|}{$m_{zAp}$} &
  Host magnitude in $z$ band \\
\multicolumn{1}{c|}{$m_{yAp}$} &
  Host magnitude in $y$ band \\
\multicolumn{1}{c|}{$\Delta m_{host, g-r}$} &
  Host magnitude subtraction: $m_{gAp}$ - $m_{rAp}$ \\
\multicolumn{1}{c|}{$\Delta m_{host, r-i}$} &
  Host magnitude subtraction: $m_{rAp}$ - $m_{iAp}$ \\
\multicolumn{1}{c|}{\textcolor{olive}{$\Delta m_{host, i-z}$}} &
  Host magnitude subtraction: $m_{iAp}$ - $m_{zAp}$ \\
\multicolumn{1}{c|}{\textcolor{olive}{$\Delta m_{host, z-y}$}} &
  Host magnitude subtraction: $m_{zAp}$ - $m_{yAp}$ \\
\multicolumn{1}{c|}{\textcolor{olive}{$\Delta \textit{colour}_{host-transient}$}} &
  $\Delta m_{host-peak, g}$ - $\Delta m_{host-peak, r}$ \\ \hline
\end{tabular}
}
\caption{\texttt{NEEDLE} upgraded metadata inputs. Features with \textcolor{olive}{olive} colour are the new ones subtracted from the old.}
\label{tab:updated-metadata}
\end{table}

\subsection{Real-time performance of NEEDLE}
Since Feb. 2024, \texttt{NEEDLE} has been running on a remote server connecting to the \texttt{Lasair} alert broker \citep{Williams_2024}. Whenever new ZTF transient alerts come into the \texttt{Lasair} database, \texttt{NEEDLE} will give the prediction on the day. Three main SQL-based filters\footnote{\href{https://lasair-ztf.lsst.ac.uk/filters/1318/run/}{SLSN candidates}; \href{https://lasair-ztf.lsst.ac.uk/filters/1330/run/}{TDE candidates}; \href{https://lasair-ztf.lsst.ac.uk/filters/944/run/}{Sources annotated in last three months}} are created to assist human scanners with assessing the candidates ranked as likely SLSNe or TDEs by \texttt{NEEDLE}. After testing and upgrading, from August 2024, the \texttt{NEEDLE} team has been publishing weekly AstroNotes via the Transient Name Server (TNS)\footnote{\href{https://www.wis-tns.org/astronotes?&posted_period_value=&posted_period_units=days&date_start=&date_end=&title=NEEDLE&}{\texttt{NEEDLE} AstroNotes}} to encourage spectroscopic follow-up of the best candidates. This follows a similar approach to community transient discovery and classification used by groups such as \texttt{ELEPHANT} \citep{Pessi_2024}, \texttt{ZTF Superluminous Supernova Science Program} \citep{Lunnan_2020, Yan_2020, Chen_2023a, Chen_2023b} and \texttt{tdescore} \citep{Stein_2024}. 

Table \ref{tab:needle_discovery} lists the SLSN-I, SLSN-II, and TDE objects that were first promoted by \texttt{NEEDLE}, and later confirmed by spectroscopic follow-up. We highlight 
{TDE 2025ccg}, which is an unusually luminous and unusually fast-evolving TDE-He, with absolute magnitude M=-21.2+/-0.2 at redshift 0.200. We are conducting ongoing follow-up of this event.

\begin{sidewaystable*}
\begin{center}
\small 
\renewcommand{\arraystretch}{1.0} 
\begin{tabularx}{\linewidth}{@{} c c c p{1.0cm} c >{\raggedright\arraybackslash}X c c c @{}}
\toprule
ZTF Object & IAU Name & Type & Redshift & Annotation Date & Detection & Probability & Spectrum Date & AstroNote \\
\midrule
ZTF25ablxhsq & 2025vjw & TDE-H-He & 0.21 & 2025-09-26 & $\mathrm{g}=23$, $\mathrm{r}=22$, $+33\,\mathrm{d}$  & 0.85 & 2025-10-24 & 2025-292 \\
ZTF25abzyzac & 2025abvv & SLSN-I & 0.19 & 2025-11-07 &$\mathrm{g}=8$, $\mathrm{r}=4$, $+14\,\mathrm{d}$  & 0.59 & 2025-11-09 & 2025-314 \\
ZTF25aaymnha & 2025qhf  & TDE      & 0.08  & 2025-07-10 & $\mathrm{g}=9$, $\mathrm{r}=6$, $+12\,\mathrm{d}$ & 0.94\rlap{$^*$} & 2025-07-19 & 2025-223 \\
ZTF25aaomnws & 2025inr  & SLSN-I/SN & 0.176 & 2025-05-25 & $\mathrm{g}=11$, $\mathrm{r}=7$, $+32\,\mathrm{d}$ & 0.72 & 2025-06-09 & 2025-175 \\
ZTF25aanikpq & 2025idz  & SLSN-I   & 0.145 & 2025-05-15 & $\mathrm{g}=6$, $\mathrm{r}=4$, $+24\,\mathrm{d}$  & 0.85 & 2025-06-01 & 2025-149 \\
ZTF25aanxtou & 2025izy  & SLSN-I   & 0.173 & 2025-04-30 & $\mathrm{g}=1$, $\mathrm{r}=2$, $+9\,\mathrm{d}$   & 0.96 & 2025-06-22 & 2025-143/161 \\
ZTF25aajbcmc & 2025esr  & SLSN-I   & 0.077 & 2025-04-11 & $\mathrm{g}=7$, $\mathrm{r}=9$, $+25\,\mathrm{d}$  & 0.54 & 2025-04-27 & 2025-119 \\
ZTF25aajjeon & 2025hbw  & TDE      & 0.628 & 2025-04-11 & $\mathrm{g}=12$, $\mathrm{r}=4$, $+21\,\mathrm{d}$ & 0.69 & 2024-04-20 & 2025-111 \\
ZTF25aaixrfr & 2025eqe  & SLSN-II  & 0.127 & 2025-04-11 & $\mathrm{g}=7$, $\mathrm{r}=6$, $+32\,\mathrm{d}$  & 0.74 & 2025-04-16 & 2025-105 \\
ZTF25aajfjhm & 2025eve  & TDE      & 0.089 & 2025-03-25 & $\mathrm{g}=4$, $\mathrm{r}=1$, $+6\,\mathrm{d}$   & 0.92 & 2025-03-26 & 2025-102 \\
ZTF25aafofcs & 2025bri  & TDE-He   & 0.061 & 2025-02-22 & $\mathrm{g}=2$, $\mathrm{r}=2$, $+7\,\mathrm{d}$   & 0.76 & 2025-02-22 & 2025-67 \\
ZTF25aafywpr & 2025ccg  & TDE-He   & 0.200 & 2025-02-20 & $\mathrm{g}=2$, $\mathrm{r}=3$, $+2\,\mathrm{d}$   & 0.86\rlap{$^*$} & 2025-02-28 & 2025-77 \\
ZTF25aagevje & 2025chm  & TDE      & 0.101 & 2025-02-24 & $\mathrm{g}=3$, $\mathrm{r}=1$, $+4\,\mathrm{d}$   & 0.92 & 2025-02-27 & 2025-71 \\
ZTF25aaaecsu & 2024agkj & SLSN-I   & 0.193 & 2025-01-17 & $\mathrm{g}=5$, $\mathrm{r}=5$, $+14\,\mathrm{d}$  & 0.78 & 2025-02-14 & 2025-32 \\
ZTF24abvzgqt & 2024adpb & SLSN-I   & 0.347 & 2024-12-11 & $\mathrm{g}=5$, $\mathrm{r}=3$, $+9\,\mathrm{d}$   & 0.62 & 2024-12-29 & 2024-368 \\
ZTF24abvftmi & 2024adtg & SLSN-II  & 0.320 & 2024-12-09 & $\mathrm{g}=5$, $\mathrm{r}=5$, $+13\,\mathrm{d}$  & 0.68 & 2024-12-27 & 2024-368 \\
ZTF24abteeyt & 2024abre & SLSN-I   & 0.170 & 2024-12-04 & $\mathrm{g}=5$, $\mathrm{r}=2$, $+14\,\mathrm{d}$  & 0.74 & 2024-12-28 & 2024-358 \\
ZTF24abtmueg & 2024adhq & SLSN-I   & 0.109 & 2024-12-05 & $\mathrm{g}=3$, $\mathrm{r}=3$, $+30\,\mathrm{d}$  & 0.94 & 2024-12-27 & 2024-358 \\
ZTF24abmzcpa & 2024zau  & TDE      & 0.082 & 2024-11-01 & $\mathrm{g}=11$, $\mathrm{r}=9$, $+20\,\mathrm{d}$ & 0.98\rlap{$^*$} & 2024-11-10 & 2024-317 \\
ZTF24abmybnp & 2024yqo  & TDE      & 0.062 & 2024-10-21 & $\mathrm{g}=4$, $\mathrm{r}=1$, $+7\,\mathrm{d}$   & 0.91 & 2024-11-09 & 2024-306 \\
ZTF24abfaake & 2024ule  & TDE-H-He & 0.111 & 2024-09-21 & $\mathrm{g}=1$, $\mathrm{r}=13$, $+19\,\mathrm{d}$ & 0.76 & 2025-01-05 & 2024-289 \\
ZTF24aazlori & 2024rny  & TDE      & 0.106 & 2024-08-16 & $\mathrm{g}=8$, $\mathrm{r}=9$, $+13\,\mathrm{d}$  & 0.57 & 2024-09-02 & 2024-231 \\
ZTF24aavxdpn & 2024qxg  & SLSN-I   & 0.450 & 2024-08-07 & $\mathrm{g}=6$, $\mathrm{r}=3$, $+29\,\mathrm{d}$  & 0.65 & 2024-08-31 & 2024-217 \\
\bottomrule
\end{tabularx}
\caption{By October 2025, \texttt{NEEDLE1.0} has predicted 12 SLSNe (10 SLSNe-I, 2 SLSNe-II) and 11 TDEs at their early stages, successfully. The earliest timestamp of over-50\% confidence is 2 days after discovery, despite very few detections (only 1--3 in $g$ or $r$ band). 
$^*$Score is an outlier that would rather classify as SLSN or TDE. ZTF25aafywpr was predicted as SLSN (due to offset and blue extended host), later confirmed as TDE after \texttt{NEEDLE} follow-up. ZTF24abmzcpa was predicted as SLSN, later confirmed as TDE.}
\label{tab:needle_discovery}
\end{center}
\end{sidewaystable*}
\section{Image Restoration, masking and up-sampling}\label{section2}

Recent studies have emphasized the importance of contextual information in transient classification. For example, \citet{Bairouk_2023} argue that contextual images carry more discriminative information than light curves, while \citet{Cao_2025} propose an image preprocessing pipeline designed to enhance overall image quality for subsequent ML training. Both works highlight that the surrounding context within transient images plays a critical role in effective feature learning and must be preserved in classification workflows. This was also the primary motivation for designing the original \texttt{NEEDLE} code.

\begin{figure*} 

\centering 
\includegraphics[width=\linewidth]{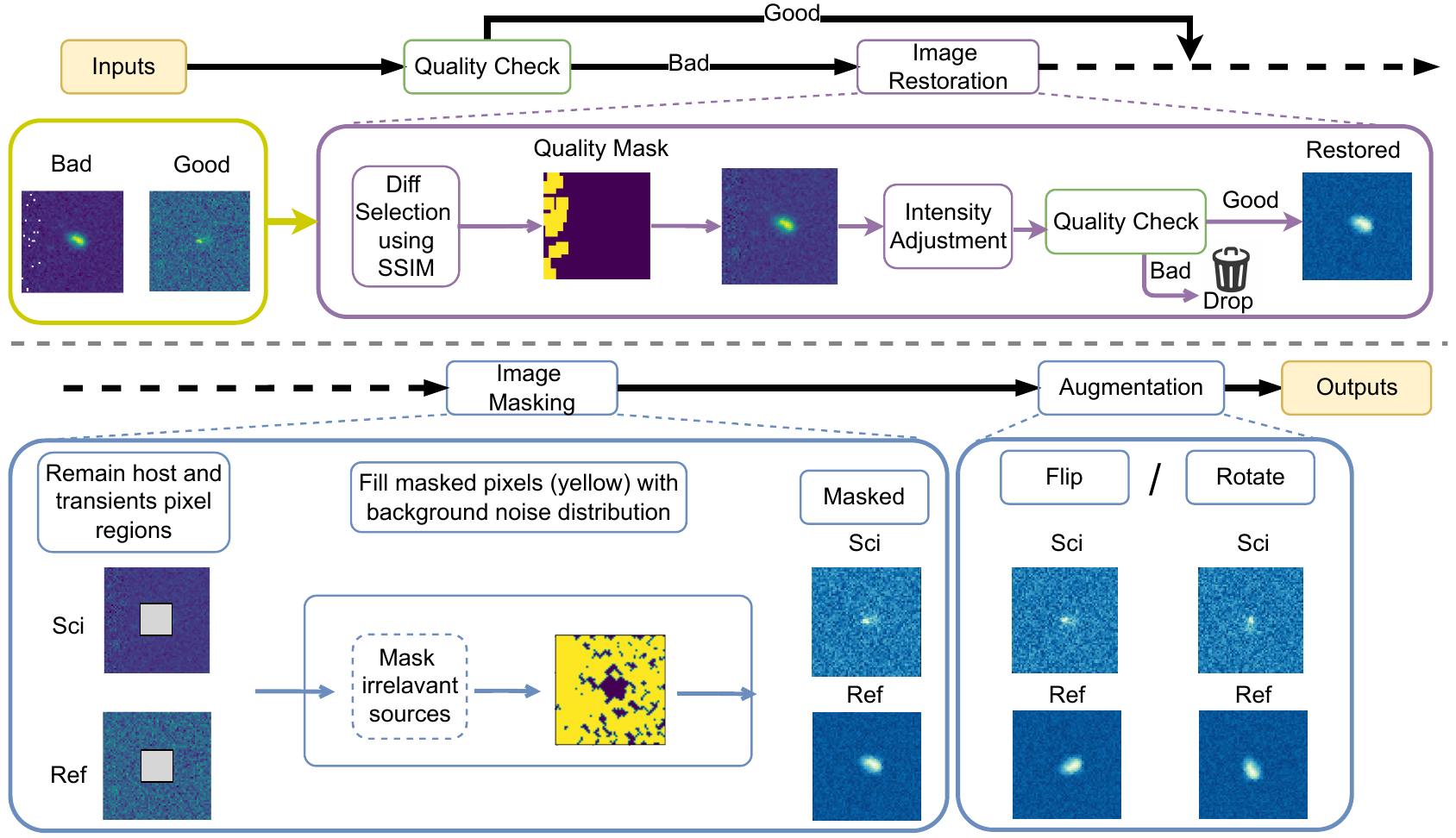} 
\caption{The NEEDLE image processing and augmentation pipeline.}
\label{fig:restoration_pipeline} 

\end{figure*}

Learning meaningful context from images is challenging when they are strongly affected by artefacts unrelated to the transient. This issue is particularly acute for rare classes, where only a limited number of examples are available. To address the dual challenges of extreme class imbalance and inconsistent image quality, we propose a set of preprocessing techniques -- comprising image restoration, attention-based masking, and augmentation strategies that generate new training samples from high-quality real observations. Importantly, our approach avoids relying on synthetic data derived from complex physical models, thereby ensuring that the augmented images remain consistent with survey characteristics. The complete image-processing pipeline is outlined in Figure~\ref{fig:restoration_pipeline} and described in detail in the following sections.

\subsection{Improved Quality Check}
\begin{figure} 
\centering 
\includegraphics[width=\linewidth]{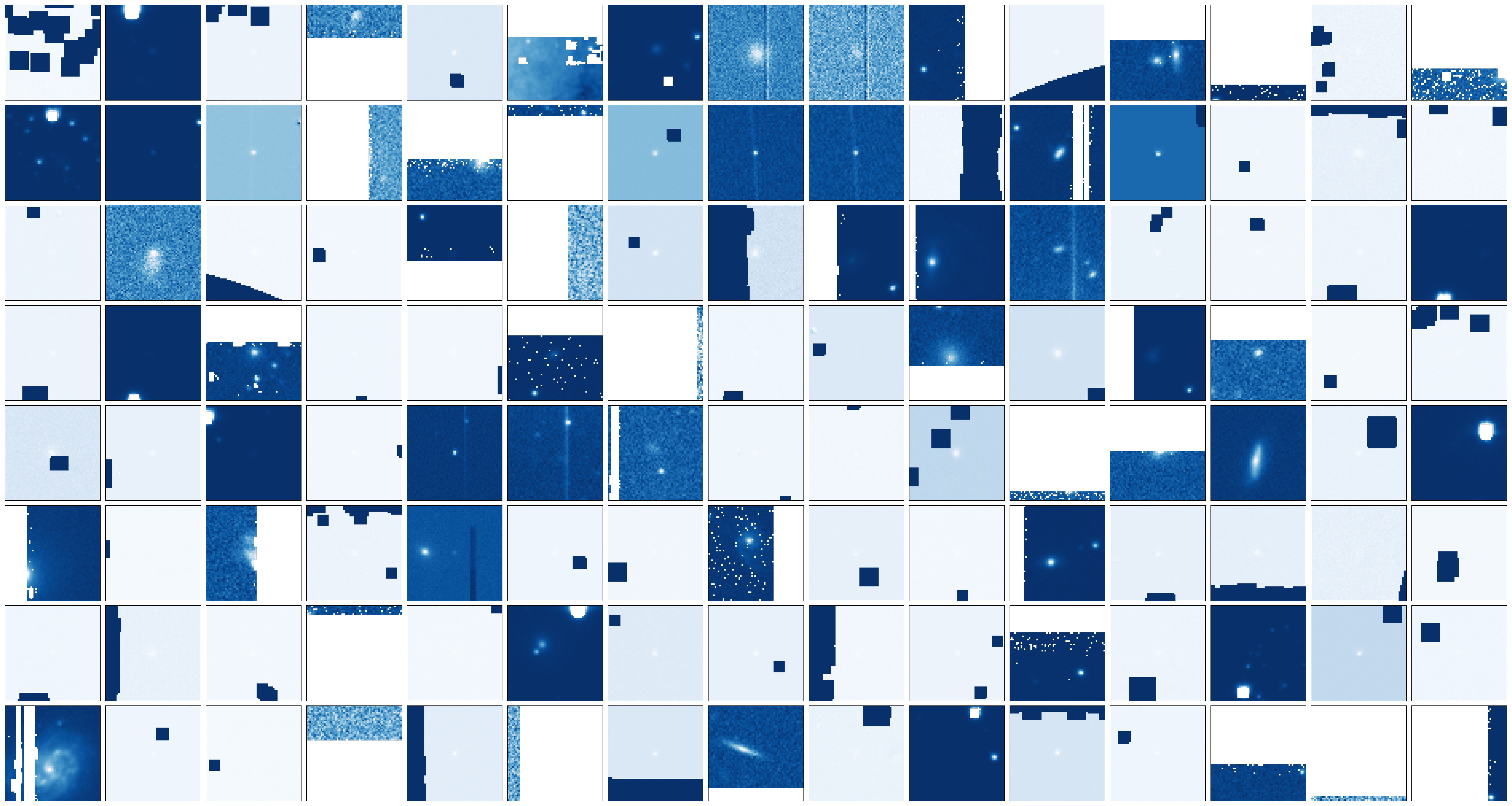} 
\caption{Examples of bad-quality images in the ZTF alert stream.} 
\label{fig:bad_images} 
\end{figure}

Although the majority of images in the ZTF data releases and alert packets are complete and informative, there are still images that suffer from data loss due to bad observing conditions, diffraction spikes from nearby stars, and transient locations close to the edge of the chip. Such images cannot be directly used for model training as they contain \textit{NaN} or anomalous values. 
In previous work \citep{Sheng_2024}, we trained a simple CNN-based classifier to give the quality score for images, using 2207 samples for training and testing, and removing images with a low quality score from later steps. While running in practice, we found that such a small training set for quality evaluation is not representative enough.
We therefore carried out a more complete search of our training set for different types of bad images, including missing pixels, bad rows/columns, extreme-high/low values, and incomplete shapes. Figure~\ref{fig:bad_images} shows the cases of bad image stamps in ZTF alerts. Table~\ref{tab:image-classifier-samples} shows the new samples for our upgraded quality checker.

\begin{table}[]
\centering
\resizebox{0.4\textwidth}{!}{%
\begin{tabular}{|c|c|c|}
\hline
Band & Good sample & Bad sample \\ \hline
$g$    & 15303       & 354        \\ \hline
$r$   & 15336       & 372        \\ \hline
Total & 30639       & 726        \\ \hline
\end{tabular}%
}
\caption{Updated sample numbers for training the image quality checker.}
\label{tab:image-classifier-samples}
\end{table}

Residual Networks (\textit{ResNet}s, \citealt{He_2016}) are a neural architecture that propagates input data through multiple layers while incorporating residual (skip) connections from earlier CNN layers. These connections help mitigate vanishing gradients and enhance feature learning. When the training dataset is sufficiently large, \textit{ResNet}s often achieve better performance than classical CNNs. In our study, we build both a plain CNN and a \textit{ResNet} classifier for comparison in order to determine the optimal model. The output is the probability that an input image is of good quality.

For the CNN, we employ two convolutional layers with 128 and 256 channels, respectively, followed by two max-pooling layers for feature downsampling. The resulting feature maps are flattened and passed through three fully connected layers, each followed by a \textit{ReLU} activation to introduce non-linearity. Finally, a \textit{log-softmax} layer produces the classification outputs.

For the \textit{ResNet} model, we adopt the standard 18-layer \textit{ResNet} (\textit{ResNet}-18) architecture. \textit{ResNet}-18 is chosen as it provides a balance between model depth and computational efficiency: it is deep enough to capture hierarchical representations while remaining feasible to train on medium-scale datasets without excessive risk of over-fitting or prohibitive training cost. 

\begin{figure} 
\centering 
\includegraphics[width=0.6\columnwidth]{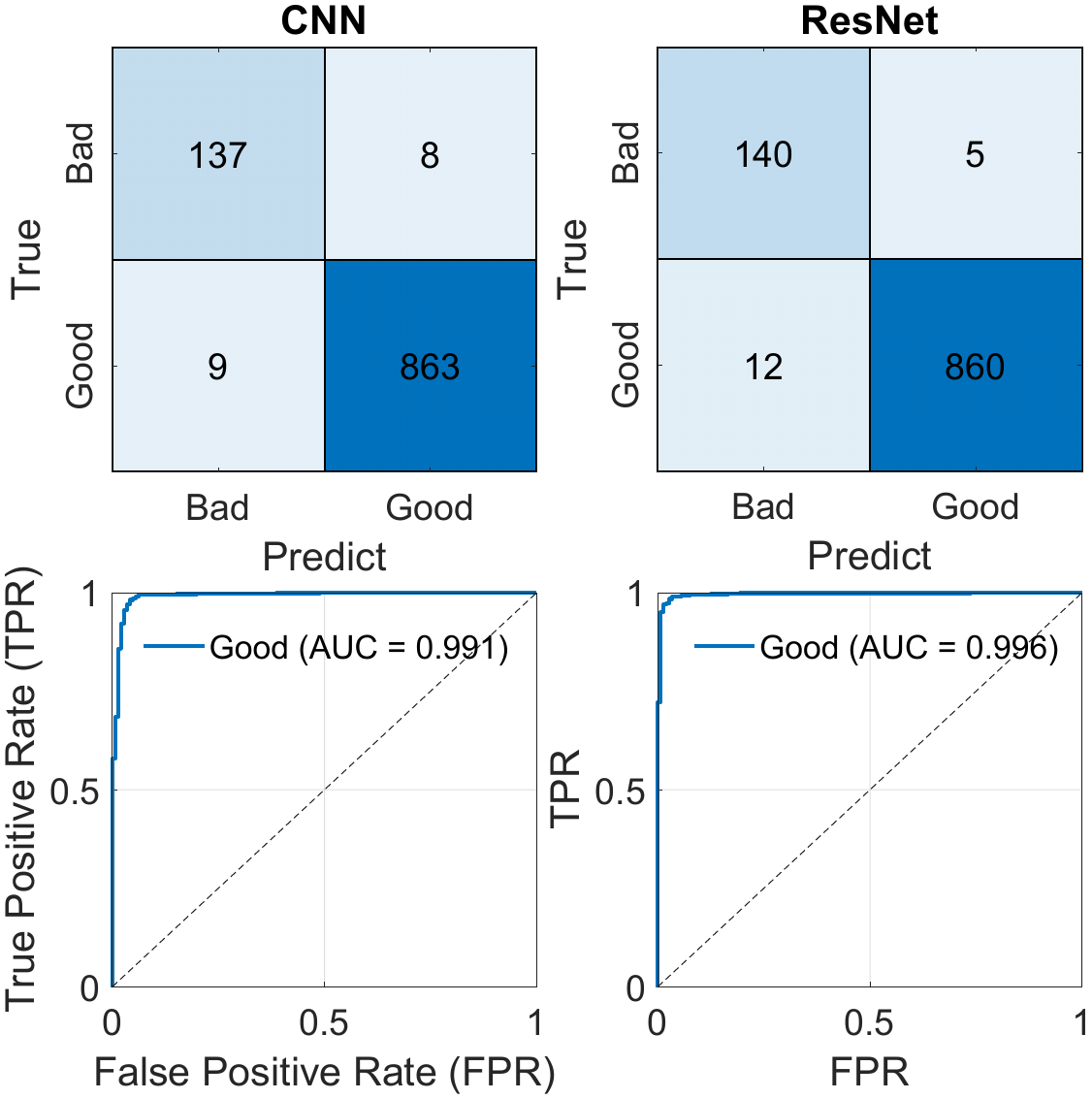} 
\caption{The comparison between CNN and \textit{ResNet} with confusion matrix and Receiver-operating characteristic curve (ROC).} 
\label{fig:CNN-ResNet-ROC}
\end{figure}

\begin{figure} 
\centering 
 \includegraphics[width=0.6\columnwidth]{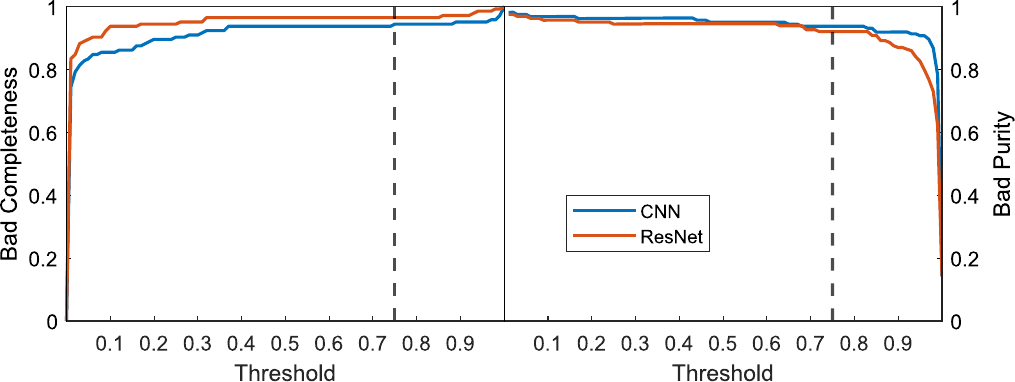} 
\caption{The comparison between CNN and \textit{ResNet} showing the completeness and purity of detecting bad images plotted against the probability output by our model. The black dashed line in completeness denotes our chosen threshold of 0.75.} 
\label{fig:CNN-ResNet-purity}
\end{figure}

Figures~\ref{fig:CNN-ResNet-ROC} and \ref{fig:CNN-ResNet-purity} show that both the plain CNN and \textit{ResNet} models perform well on the image quality classification task. \textit{ResNet} slightly outperforms the CNN by achieving higher completeness (recall) for the True Negative (bad sample) class. The relatively low purity of the True Negatives is less critical, as the predicted bad samples will be restored in the subsequent processing stage, and the False Negatives will still be retained in the training set.

\subsection{Image Restoration}

During this work, we observed two common scenarios: in the first, poor image quality occurs in only one detection out of the many observations that make up the light curve; in the second, all detections suffer from the same image quality issues. These scenarios can affect both the \textit{Science} (transient detection) and \textit{Reference} (pre-transient) images. When only a subset of images is affected, it is straightforward to iterate and select the best inputs for \texttt{NEEDLE}, e.g.\ by selecting the good image closest in time. However, if either all \textit{Science} or \textit{Reference} images are of low quality, there is still a chance to restore them, instead of discarding them entirely. We adopt the following approach:

\textbf{1. Find restoration cases.}
For any bad images, we search for high quality images of the same field. If the \textit{Science} image is of bad quality, we trace back up to five prior detections to check for any high-quality \textit{Science} images that we could use instead. If all \textit{Science} images are of poor quality, the restoration procedure will be triggered to reconstruct the \textit{Science} image using the \textit{Reference}. Conversely, if the \textit{Reference} image is of low quality but there are good \textit{Science} images available, we restore the \textit{Reference} using the \textit{Science} images.

\textbf{2. Size padding.}
Some images have incomplete sizes (smaller than the 60×60 pixels expected by our CNN). Since the transient is always designated to be located at the centre of the image, we use the information from the image \texttt{header} to determine its position and pad the missing rows or columns of the image with pixels containing an extreme value \textit{-999} (which will then be repaired in steps 4 and 5) until the image reaches the required 60×60 size.

\begin{figure*}
    \centering
    
    \includegraphics[width=\textwidth, height=0.85\textheight, keepaspectratio]{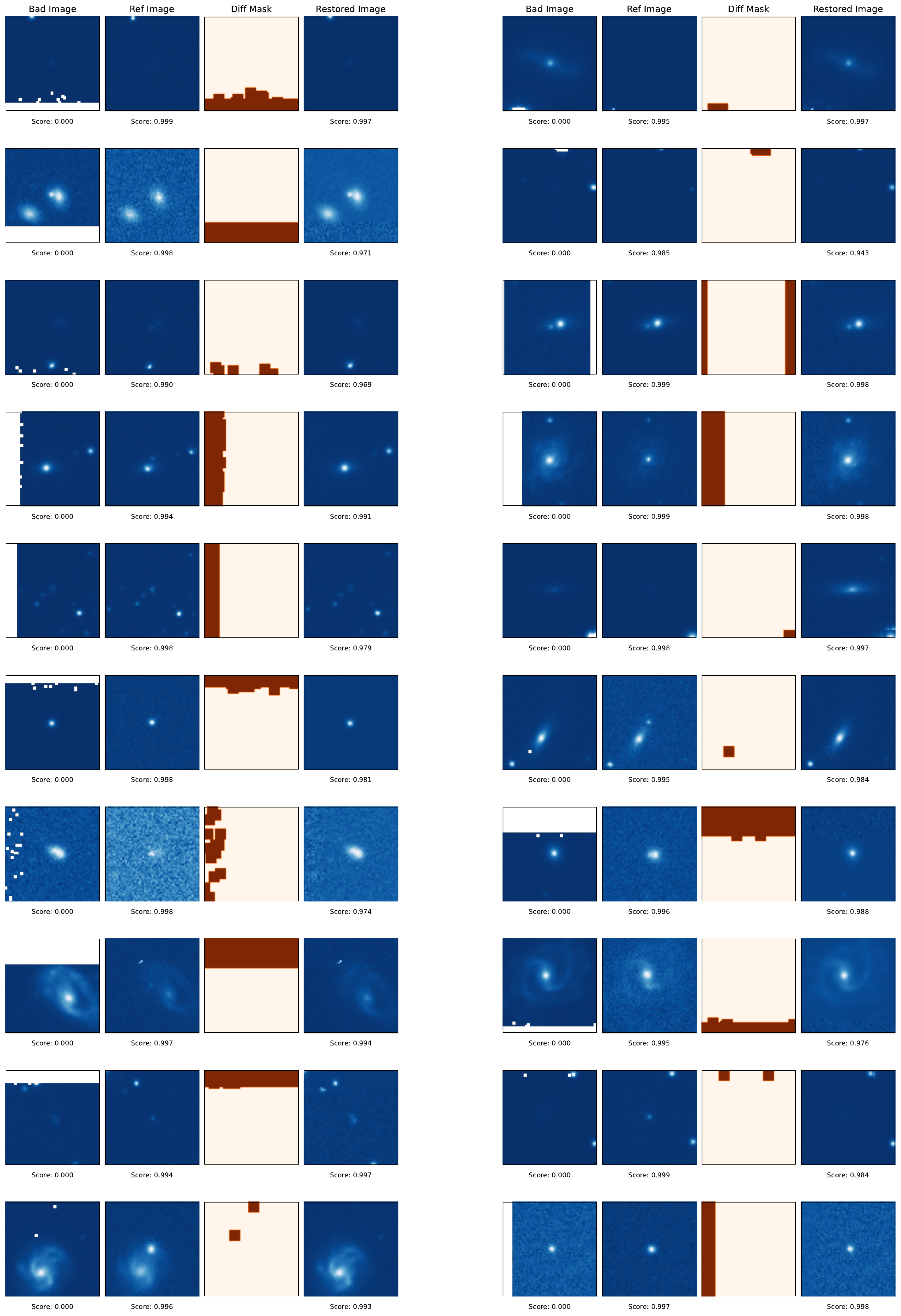}
    \caption{Image restoration samples. The score shown below each image is generated by the quality checker. From left to right: bad images, a good reference image of the same field, the mask of bad pixels calculated by comparing these images using \texttt{SSIM}, and the final restored image after restoration. In the restoration process, masked pixels in the bad image are replaced with scaled values from the reference image, ensuring consistency with the original noise distribution. All restored images achieve scores above the 0.5 threshold, confirming their classification as good quality.}
    \label{fig:restoration_samples}
\end{figure*}

\textbf{3. Identify image outliers.}
Structural Similarity Index (\texttt{SSIM}; \citealt{Zhou_2004}) is a quality assessment algorithm based on the degradation of structural information. Here we apply \texttt{SSIM} to compare the bad image (which could be a \textit{Reference} or \textit{Science} image) to a known good reference image, and identify the anomalous parts. The mask produced by the \texttt{SSIM} is shown as the `Diff Mask' in Figure~\ref{fig:restoration_samples}.

\textbf{4. Intensity adjustment.}
As images of the same field are taken at different times, the distributions of pixel values will vary. To enable substitutions of good pixels from one image for bad pixels in another, we first scale images to the same intensity distributions. The intensity adjustment for the masked area follows
\begin{align}
    {\lambda} &= \frac{\sigma_{S_b}}{\sigma_{S_g}} \nonumber \\[0.5ex]
    S_{m'} &= \bigl(S_m - \bar{S_g}\bigr) \cdot \lambda + \bar{S_b}, \label{eq:density-adjust}
\end{align}
where $S_m$ ($S_{m'}$) is the \texttt{SSIM}-masked pixel region to be restored (restored), $\bar{S_g}$ ($\bar{S_b}$) and $\sigma_{S_g}$ ($\sigma_{S_b}$) are the mean and standard deviation of pixel values in the good (bad) image, respectively (\textit{NaN} values excluded).

\textbf{5. Quality check on restored images.} We then re-run the quality checker on the restored image. If the restored image still fails, it will be removed.

Using this method, we `recycle' 91 poor-quality images from our training set. Our successful pipeline can be employed to process streaming alerts on \texttt{Lasair}, which makes sure the majority of real-time alerts will be classified through \texttt{NEEDLE}, rather than being rejected due to image artefacts.

Although we are now able to remove image artefacts, the images still contain many real sources that are not related to the transient, and can be distracting for the CNN. Moreover, the total numbers of SLSNe and TDEs in the NEEDLE training set remain small. Therefore, we extend our approach to also remove unrelated sources, which will later aid in upsampling these classes. 

\subsection{Image Masking}

In \texttt{NEEDLE1.0}, imaging inputs are passed to the CNN model after basic batch normalization and augmentation. In practice we found this process often causes the model to focus more on nearby brighter sources rather than on the transient and its host, especially for SLSNe-I with faint hosts. In this work, we improve image learning by masking out all nearby irrelevant sources while retaining only the transient and its potential host. This encourages the CNN to focus its attention on the most relevant pixels. We adopt the following procedure:

\textbf{1. Produce a sample of image background and realistic noise.} 
First, we use sigma clipping to generate a binary mask $M_b$ representing bright sources, where sigma is the standard deviation of the pixel values. Pixels below this threshold are considered part of the noise mask $M_n$, while pixels above this are considered possible astronomical sources. However, since $M_n$ may still contain extended luminous regions surrounding bright sources, an additional clipping step is applied to refine the noise mask. 

\begin{align}
\text{Convert the image matrix I into a flat array:} \quad & \text{Let } \tilde{I} = \text{flatten}(I) \nonumber \\[0.5ex]
\text{Remove NaNs and compute mean and std:} \quad & \mu = \text{mean}(\tilde{I}), \quad \sigma = \text{std}(\tilde{I}) \nonumber \\[0.5ex]
\text{Mask to separate bright and noisy pixels:} \quad & \text{M}_b = \{ x \in \tilde{I} \mid x > \mu + f_1 \cdot \sigma \}\nonumber \\
& \text{M}_n = \{ x \in \tilde{I} \mid x \leq \mu + f_1 \cdot \sigma \}, \quad f_1 = 3 \nonumber \\[0.5ex]
\text{Noise image (non-zero elements only):} \quad & I_{\text{noise}} = \{ x \in \text{M}_n \mid x \neq 0 \} \nonumber \\[0.5ex]
\text{Compute stats on the noise distribution:} \quad & \mu' = \text{mean}(I_{\text{noise}}), \quad \sigma' = \text{std}(I_{\text{noise}}) \nonumber \\[0.5ex]
\text{Select values below secondary threshold:} \quad & \text{N} = \{ x \in I_{\text{noise}} \mid x < \mu' + f_2 \cdot \sigma' \}, \quad f_2 = 1 \nonumber \\[0.5ex]
\text{Final Noise set:} \quad & N_{\text{final}} = \{ x \in \text{N} \mid x \neq \text{NaN} \} \label{eq:get-noise}
\end{align}

The parameters $f_1 = 3$ and $f_2 = 1$ (i.e. clipping at $3\sigma$ and then at $1\sigma$) were determined empirically through experimentation.

\textbf{2. Pixel clustering.} 
Here we apply Density-Based Spatial Clustering of Applications with Noise (\texttt{DBSCAN}; \citealt{Ester_1996}) for pixel clustering. It is an unsupervised machine learning algorithm used to group data points into clusters based on density, well-suited for transient images. Clusters of sources comprised of adjacent pixels are formed and labeled. It is worth noting that two sources might reside in the same cluster due to their luminosity overlap and closeness.  

\textbf{3. Find clusters containing transient and host.}
The most likely host galaxy is identified using \texttt{Sherlock} \citep{Smith_2020}. We use the \texttt{header} information to obtain the pixel coordinates of the transient and its host, and identify the corresponding clusters they reside in. The \textit{Science} and \textit{Reference} images are processed separately to generate individual masks, $M_{\text{sci}}$ and $M_{\text{ref}}$, respectively. 
Since some faint transient hosts may not have been catalogued previously but could still be viable, we define an additional fixed mask, $M_{\text{fixed}}$, as a 3×3 pixel region centred on the image. The final mask, $M_{\text{final}}$, is then constructed as the union of the three masks:

\begin{equation}
    M_{\text{final}} = M_{sci} \cup M_{Ref} \cup M_{fixed}. 
\end{equation}

This ensures that any faint (detected only in the deeper Reference image) or very compact hosts (which may be missed by \texttt{DBSCAN}) are included in the mask.

\textbf{4. Create the mask for irrelevant nearby sources.}
By definition, the inverse mask represents the mask of irrelevant pixels. These regions are replaced with random noise values sampled and shuffled from $N_{\text{final}}$.

\textbf{5. Density adjustment between $Science$ and $Reference$ images.} 
Finally, to reduce the effects of absolute pixel values on image learning, we use the same Equations~\ref{eq:density-adjust} to scale the two images used for classification to identical $\mu$ and $\sigma$, where $S_x$ is the Science image, and $S_y$, $S_z$, and $S_{z'}$ are replaced by the Reference image.

\begin{figure*}
    \centering
    \includegraphics[width=\textwidth, height=0.85\textheight, keepaspectratio]{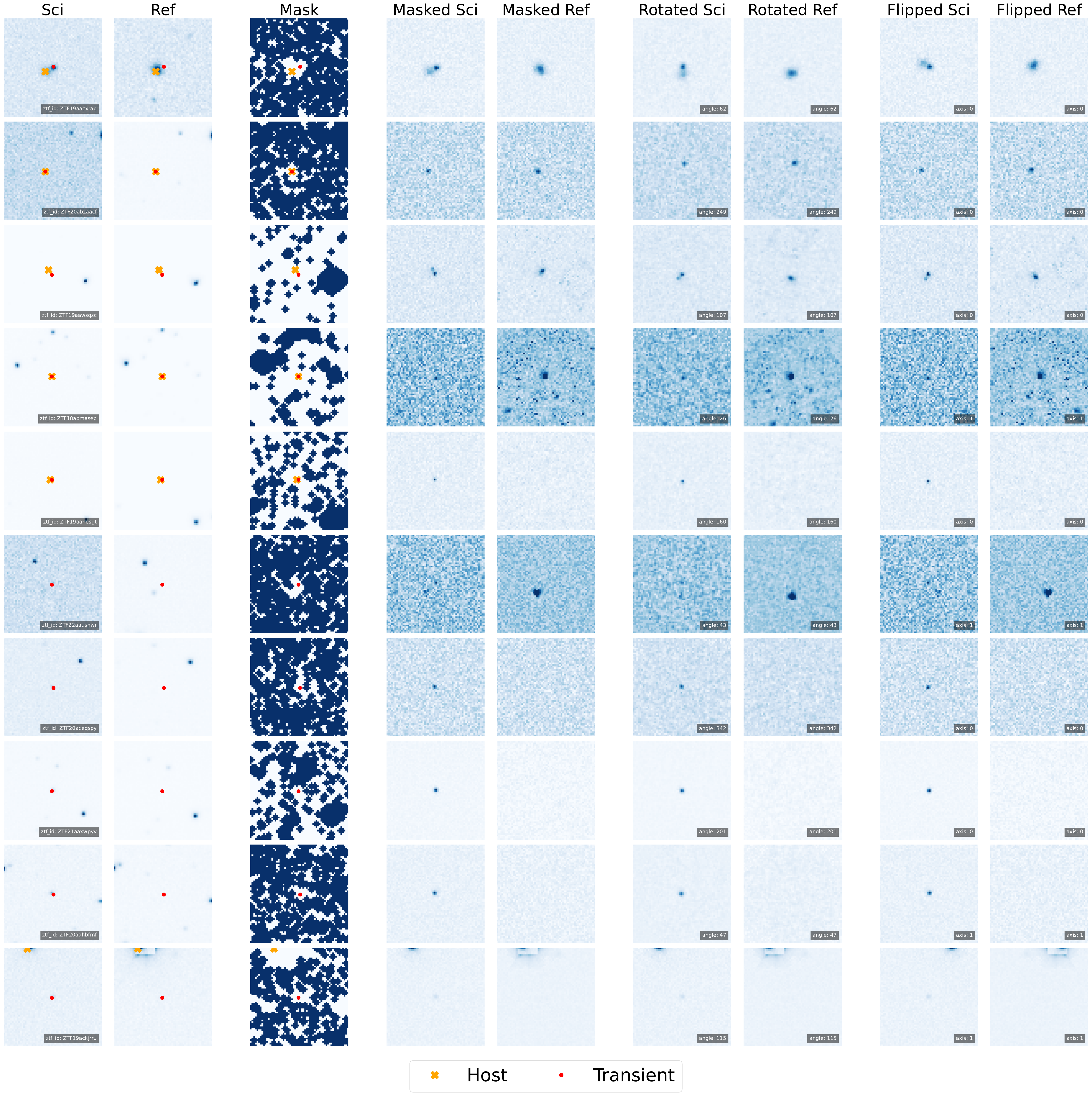}
    \caption{SLSN-I image augmentation. The first two columns show the original \textit{Science} (Sci) and \textit{Reference} (Ref) images. The third column presents the boolean mask used to identify and remove irrelevant sources; the host galaxy (marked with an orange cross, when available) and the transient (marked in red) are retained. The fourth and fifth columns display the masked \textit{Science} and \textit{Reference} images. The remaining columns show the augmented products after applying arbitrary rotations and flips, which are used for upsampling.}
    \label{fig:slsn_mask}
\end{figure*}

\begin{figure*}
    \centering
    \includegraphics[width=\textwidth, height=0.85\textheight, keepaspectratio]{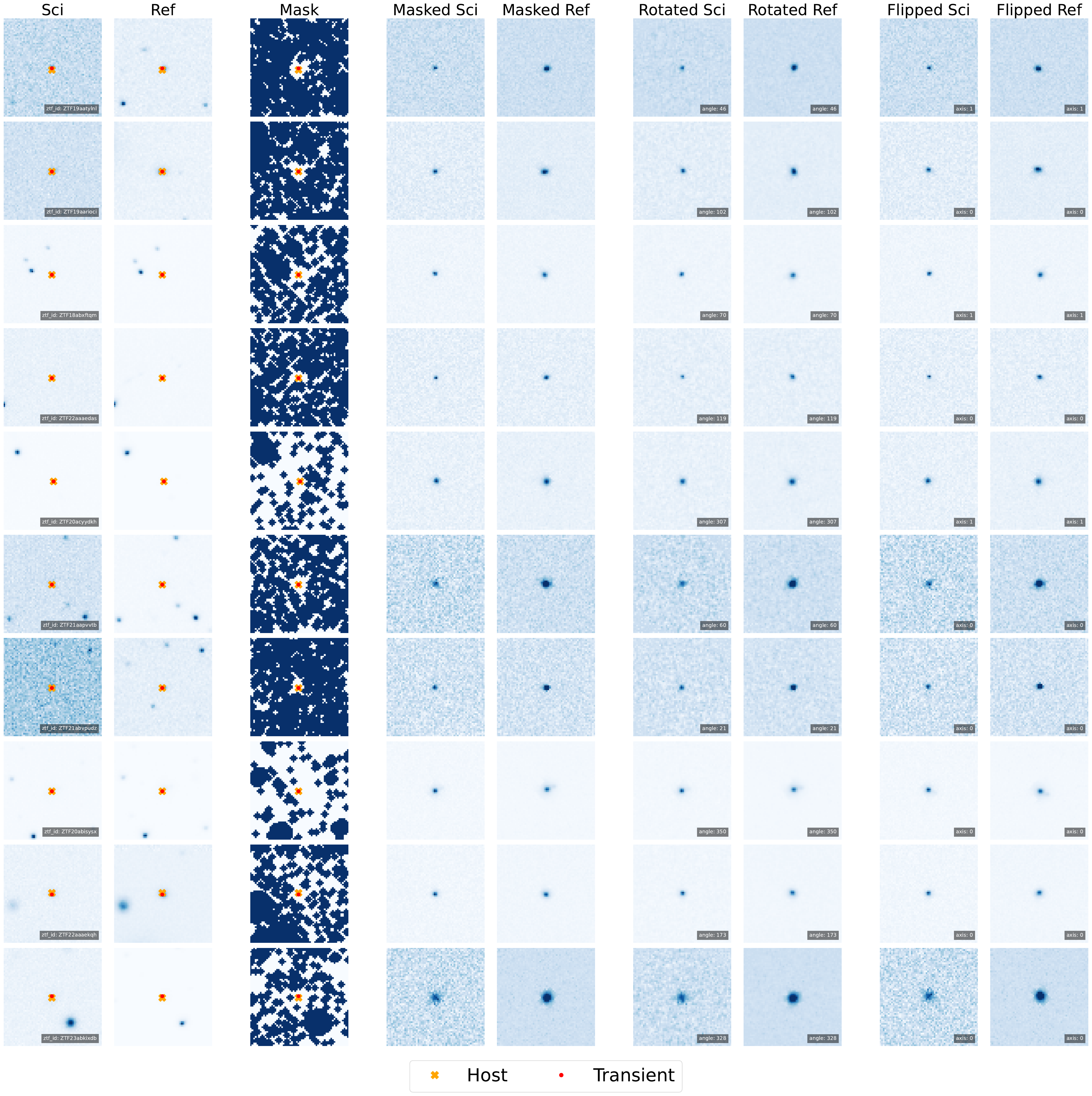}
    \caption{TDE image augmentation. The details are the same as in Figure~\ref{fig:slsn_mask}, but applied to the TDE sample.
}
    \label{fig:tde_mask}
\end{figure*}

\begin{figure*}
    \centering
    \includegraphics[width=\textwidth, height=0.85\textheight, keepaspectratio]{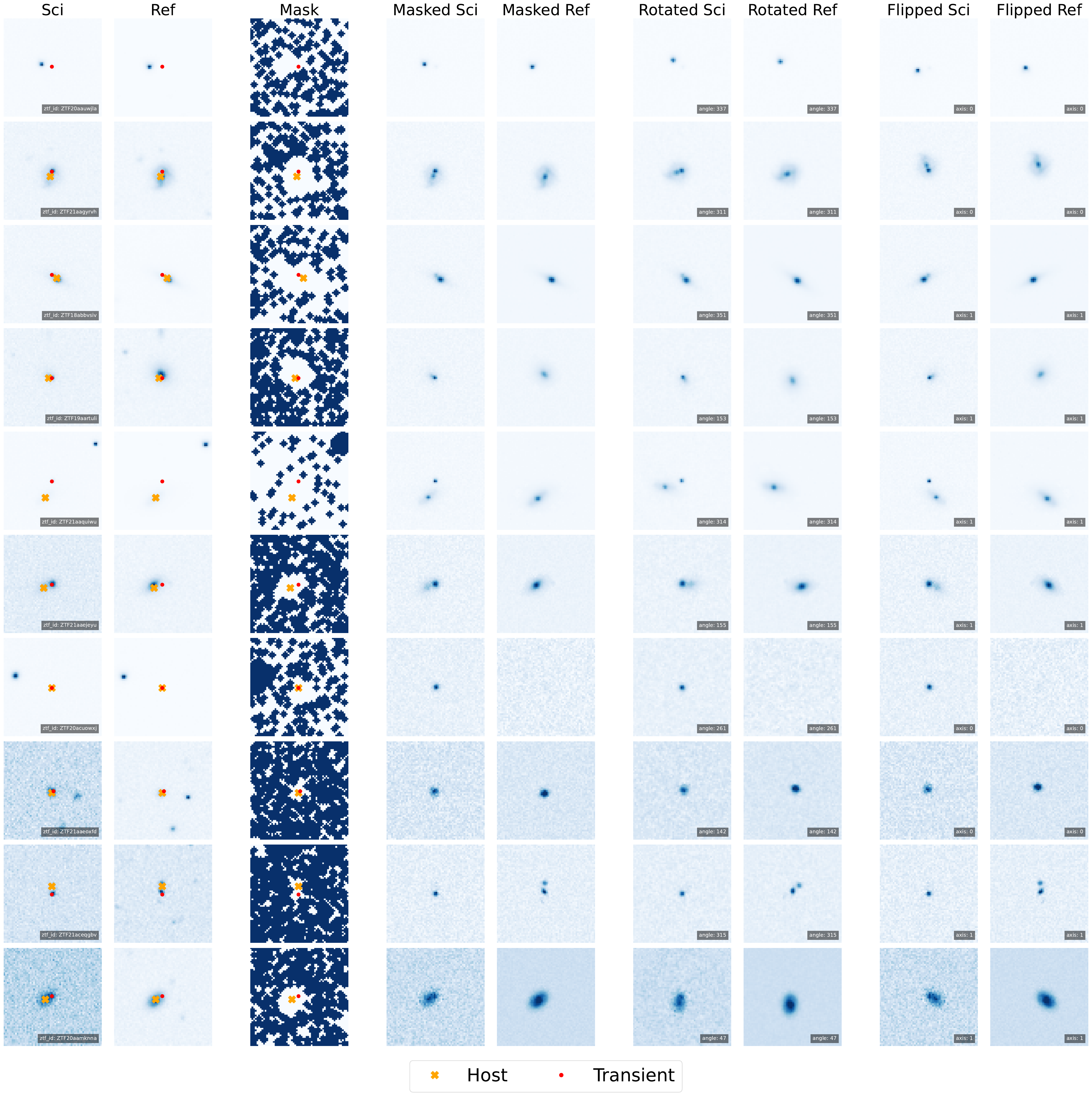}
    \caption{SN Ia image augmentation. The details are the same as in Figure~\ref{fig:slsn_mask}, but applied to the SN Ia sample.
}
    \label{fig:ia_mask}
\end{figure*}

\begin{figure*}
    \centering
    \includegraphics[width=\textwidth, height=0.85\textheight, keepaspectratio]{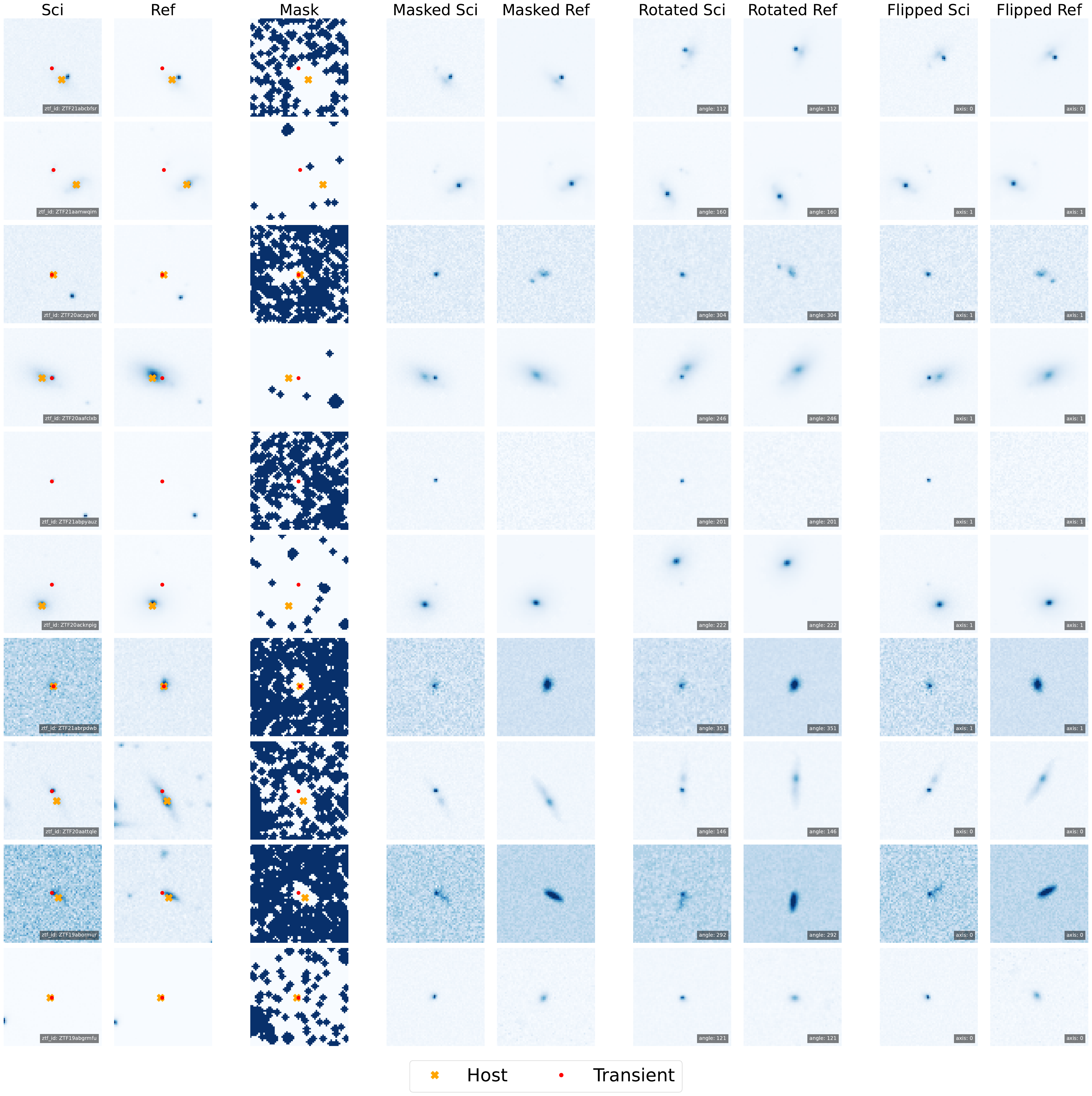}
    \caption{SN II image augmentation. The details are the same as in Figure~\ref{fig:slsn_mask}, but applied to the SN II sample.}
    \label{fig:ii_mask}
\end{figure*}

The first five columns in Figures~\ref{fig:slsn_mask}, \ref{fig:tde_mask}, \ref{fig:ii_mask}, and \ref{fig:ia_mask} show examples of the masking process for the SLSN-I, TDE, SN II, and SN Ia classes. The latter two are grouped under the SN label in both \texttt{NEEDLE1.0} and \texttt{2.0}. These examples demonstrate the effectiveness of the masking strategy--it successfully removes most nearby irrelevant sources and compels the model to \textbf{pay all attention} to the pixels corresponding to the transient and its host.

\subsection{Image Augmentation}
The final stage of image preprocessing is augmentation.  
It prevents the network from attaching any meaning to the random orientation and spatial distributions of celestial objects, while also enabling the generation of additional samples to help upsample the rare classes.

In \texttt{NEEDLE1.0}, this was implemented using simple operations: rotations by 90°, 180°, or 270°, and random flips along both axes. In our current approach, since all nearby sources are masked out, images containing a complete view of the transient and its host can be rotated or flipped by \textbf{any} degree. Any interpolated empty regions resulting from these transformations are filled with values sampled from $N_{\text{final}}$. If the host is incomplete because it extends beyond the edge of the original image cutout, the image is augmented using the original method (limited to 90°, 180°, or 270° rotations and flips) to avoid distortion.

\subsection{Attention-Guided CNN Heatmaps for Masked Images}

\begin{figure}
    \centering
    \includegraphics[width=0.65\linewidth]{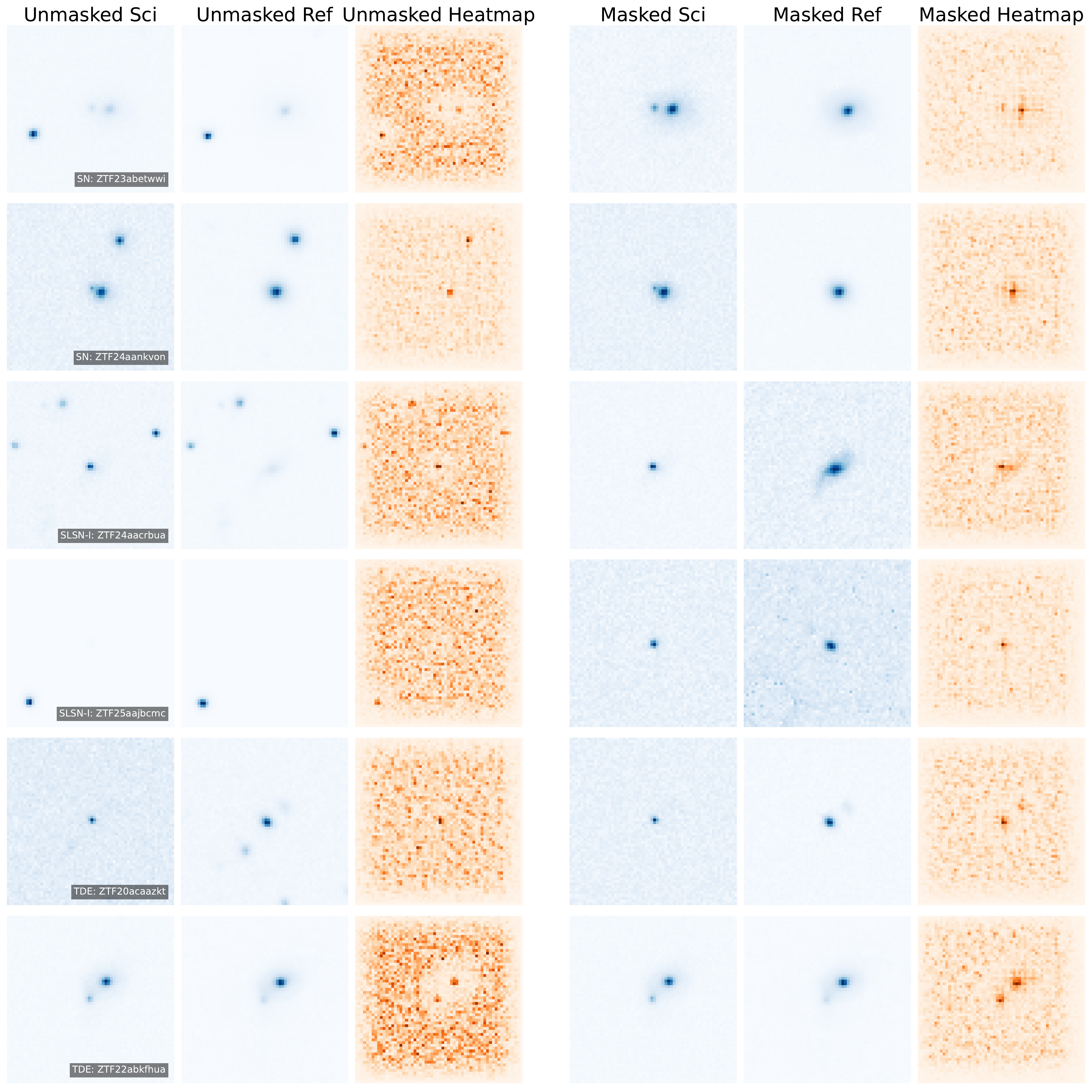}
    \caption{Heatmap comparisons before and after adding the Masking procedure. The upper, middle, and bottom two columns are selected SN, SLSN-I and TDE samples.
    In the heatmap columns, darker colours correspond to higher attention.}
    \label{fig:Heatmap}
\end{figure}

To demonstrate that the masking procedure facilitates the ability of the CNN to concentrate on the transient and host galaxy, we extract the activations from the final convolutional layer and generate class activation heatmaps before and after training on both the original and masked image sets. The corresponding heatmap comparisons are presented in Figure \ref{fig:Heatmap}. Supplying the raw, unmasked images to the CNN can cause the network to attend simultaneously to both relevant and irrelevant sources, thereby diluting the contribution of the truly informative context. This effect is particularly pronounced for SLSN-I images, as their host galaxies are typically much fainter than nearby background objects. Once all neighboring stars and galaxies are removed by the masking process, the resulting heatmaps show that the network’s attention is predominantly concentrated on the appropriate pixels.

\section{Light curve re-sampling}\label{section3}

Since \texttt{NEEDLE} takes both images and light curves as inputs, each additional sample generated through image augmentation must be paired with a corresponding light curve. Reusing identical light-curve measurements for these augmented images would likely introduce overfitting. Moreover, variations in the number of detections in original light curves may introduce systematic biases when learning correlations between image-based and photometric features. Consequently, to enrich the labeled dataset and fully exploit ensemble information from the complete sample, we need a method to generate a large set of realistic light curves that can be flexibly cross-matched with arbitrarily augmented images.

Here we present a light curve pipeline that includes two-dimensional Gaussian Process (2D GP) modelling and methods that use these fits to produce new light curve samples, at different redshifts or observed with different cadence. 
This procedure is generalizable and can be applied to any light curve data. 
Since the primary goal of \texttt{NEEDLE} is to classify infant transients, only data up to and around the light curve peak (latest time is $\text{MJD}_{peak}+5\text{d}$) are simulated.

\subsection{Light curve pre-processing}
 
We start with the real observed light curves from ZTF in the $g$ and $r$ bands. 
First, dust extinction is removed using \texttt{PyPI} packages \href{https://extinction.readthedocs.io/en/latest/}{\texttt{extinction}} and \href{https://pypi.org/project/extinctions/}{\texttt{extinctions}} with \texttt{SFD} dust maps \citep{Schlafly_Finkbeiner_2011}). These packages query the position of the transient to determine the Milky Way reddening in that direction, then calculate the extinction in magnitudes in each band. These are then subtracted from the ZTF light curves. 
We neglect host galaxy extinction, as there is no straightforward way to estimate this for such a large and diverse transient sample. It is not a problem for generating simulated light curves, since the simulated transients will then have the same (unknown) extinction as the set of examples used to generate them. 

We then merge detections within a 0.5-day time window and compute the average magnitude in each band within that window. This step is necessary because \texttt{NEEDLE} features include the recent light curve slope. This is highly sensitive to short time gaps: random photometric scatter can blow up into large apparent slopes when dividing by short time intervals.

\subsection{Two-dimensional GP}

Two-dimensional Gaussian Process (2D GP) fitting \citep{Boone_2019} employs a kernel that spans both the wavelength and time dimensions, enabling the model to capture correlations across photometric bands and over time. 
This approach is particularly effective for interpolating missing observations (e.g.\ if a gap exists in the light curve of one band, the GP model can still be constrained by the other band) and is well-suited to datasets with irregular sampling across multiple epochs and filters.

In this work, we intentionally avoid assuming any physical transient models in order to minimize bias. 
The physics of SLSN and TDE light curves are not fully understood, and even for normal SNe no simple model will capture the full diversity and complexity observed in nature.
Following the approach in \citet{Boone_2019}, we adopt a \texttt{Matérn-3/2 covariance kernel} to learn the underlying covariance structure directly from the observed data. This method holds promise as a useful tool for analysing ZTF and LSST transients, demonstrating its ability to learn correlations across all six bands. 
While 2D GP is powerful in modelling the temporal and spectral correlations of individual transients, its effectiveness depends on the number of available observations. In addition, GP fitting is not always consistent with the underlying physical processes, especially in non-stationary systems \citep{Stevance_2023}. 
At the very least, it provides a useful benchmark for illustrating the evolution of inter-band relationships.

We restrict the GP modelling to objects with more than two detections in $g$ or $r$ band during the rising phase (but one detection at least at the up-sampling step).
We apply the \texttt{UnivariateSpline} method to smooth the light curve and identify outliers. The principle is that the spline provides a smooth, continuous approximation to the data; points that deviate significantly (here, by more than three standard deviations from the fitted spline) are considered outliers and are removed. 
In addition, to avoid influence from any spurious, pre-transient flux (e.g.~imperfect image subtraction residuals, or unrelated AGN variability in the same galaxy), we discard data points that precede a detection gap larger than 100 days. 

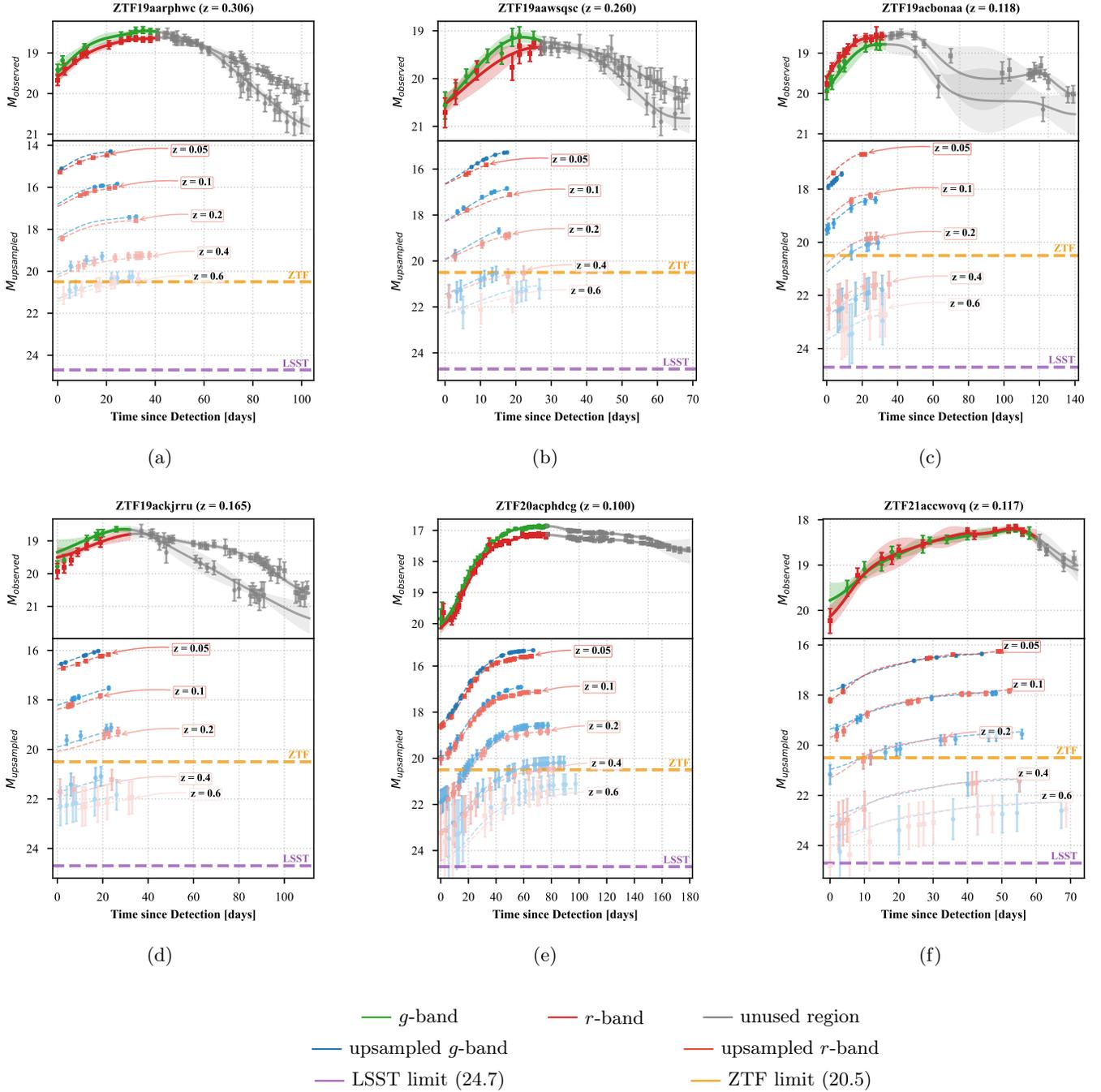
\begin{figure*}[htbp]
    \centering

    \begin{subfigure}[b]{0.3\textwidth}
    \includegraphics[width=\textwidth]{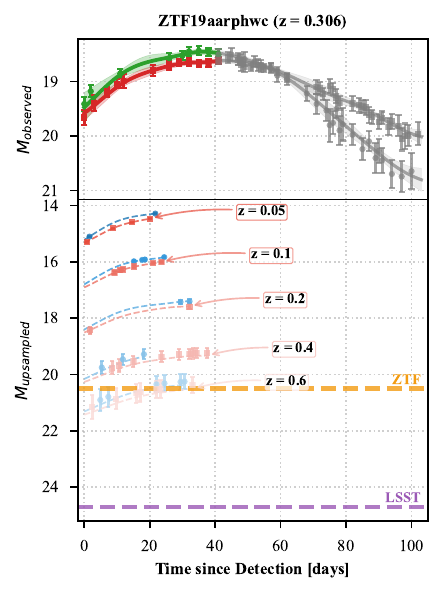}
        \caption{}
    \end{subfigure}
    \hfill
    \begin{subfigure}[b]{0.3\textwidth}
        \includegraphics[width=\textwidth]{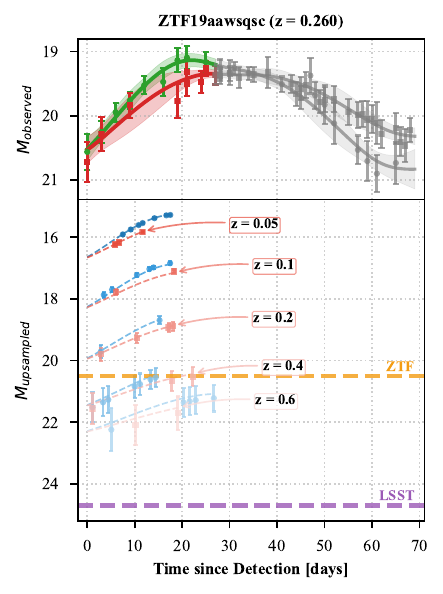}
        \caption{}
    \end{subfigure}
    \hfill
    \begin{subfigure}[b]{0.3\textwidth}
        \includegraphics[width=\textwidth]{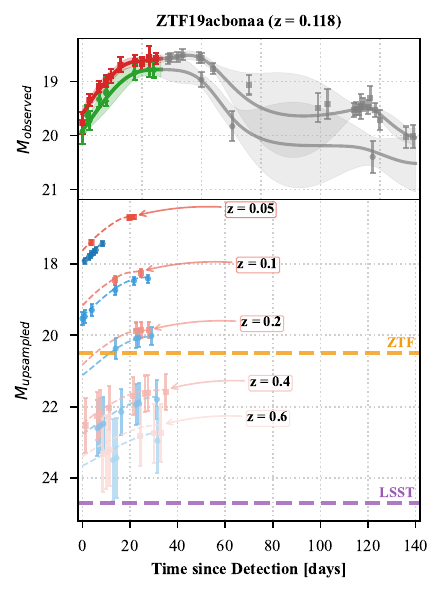}
        \caption{}
        \label{fig:slsn-bag-gp}
    \end{subfigure}

    \vspace{4pt}
    \begin{subfigure}[b]{0.3\textwidth}
        \includegraphics[width=\textwidth]{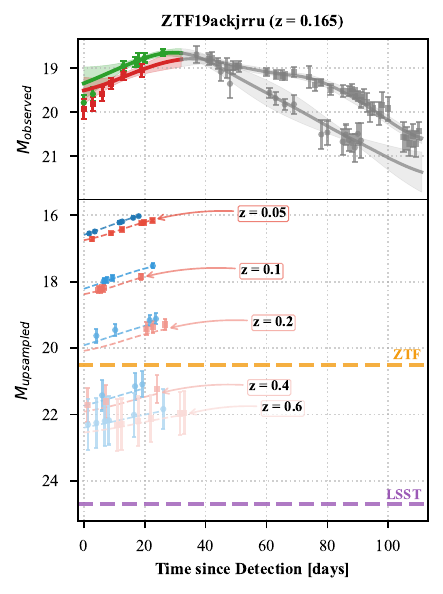}
        \caption{}
    \end{subfigure}
    \hfill
    \begin{subfigure}[b]{0.3\textwidth}
        \includegraphics[width=\textwidth]{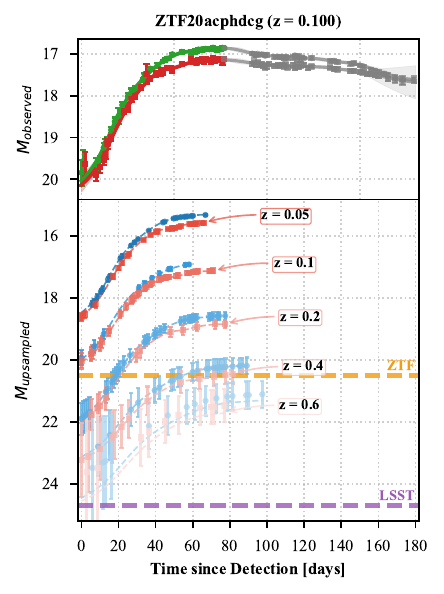}
        \caption{}
    \end{subfigure}
    \hfill
    \begin{subfigure}[b]{0.3\textwidth}
        \includegraphics[width=\textwidth]{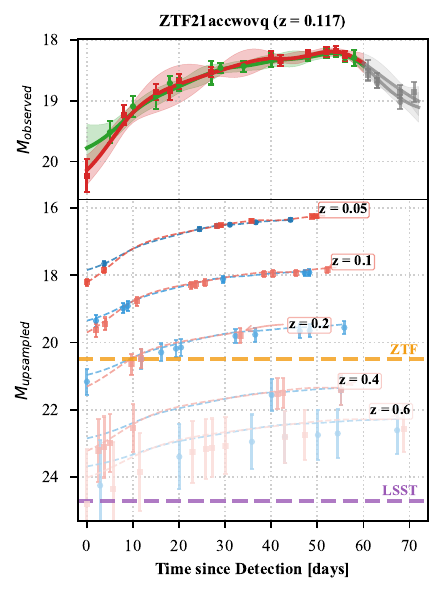}
        \caption{}
    \end{subfigure}

    \vspace{10pt}
    \centering
    \begin{tabular}{@{}c@{\hspace{15pt}}c@{\hspace{15pt}}c@{\hspace{15pt}}c@{\hspace{15pt}}c@{}}
   
        \begin{tikzpicture}[baseline=-0.5ex]
            \draw[myGreen, thick] (0,0) -- (1.5em,0) node[right, black] {$g$-band};
        \end{tikzpicture} &
        \begin{tikzpicture}[baseline=-0.5ex]
            \draw[myMaroon, thick] (0,0) -- (1.5em,0) node[right, black] {$r$-band};
        \end{tikzpicture} &
        
        \begin{tikzpicture}[baseline=-0.5ex]
            \draw[gray, thick] (0,0) -- (1.5em,0) node[right, black] {unused region};
        \end{tikzpicture} \cr
        \begin{tikzpicture}[baseline=-0.5ex]
            \draw[myBlue, thick] (0,0) -- (1.5em,0) node[right, black] {upsampled $g$-band};
        \end{tikzpicture} & &
        \begin{tikzpicture}[baseline=-0.5ex]
            \draw[myRed, thick] (0,0) -- (1.5em,0) node[right, black] {upsampled $r$-band};
        \end{tikzpicture} \cr
        \begin{tikzpicture}[baseline=-0.5ex]
            \draw[myPurple, thick] (0,0) -- (1.5em,0) node[right, black] {LSST limit (24.7)};
        \end{tikzpicture} & &
        \begin{tikzpicture}[baseline=-0.5ex]
            \draw[myOrange, thick] (0,0) -- (1.5em,0) node[right, black] {ZTF limit (20.5)};
        \end{tikzpicture}
    \end{tabular}
    \caption{Six examples of resampled SLSN-I light curves. For each sample, the upper panel shows the original detections with a two-dimensional Gaussian Process fit. Only the rising phase in the $g$ and $r$ bands (shown in green and red, respectively) is used to extract photometric features. The lower panels display the resampled light curves over a wide range of redshifts $z$. At higher redshifts, detections are truncated by the ZTF observing limit, whereas the majority of detections remain observable within the LSST depth.}
    \label{fig:slsn-lc-upsample}
\end{figure*}

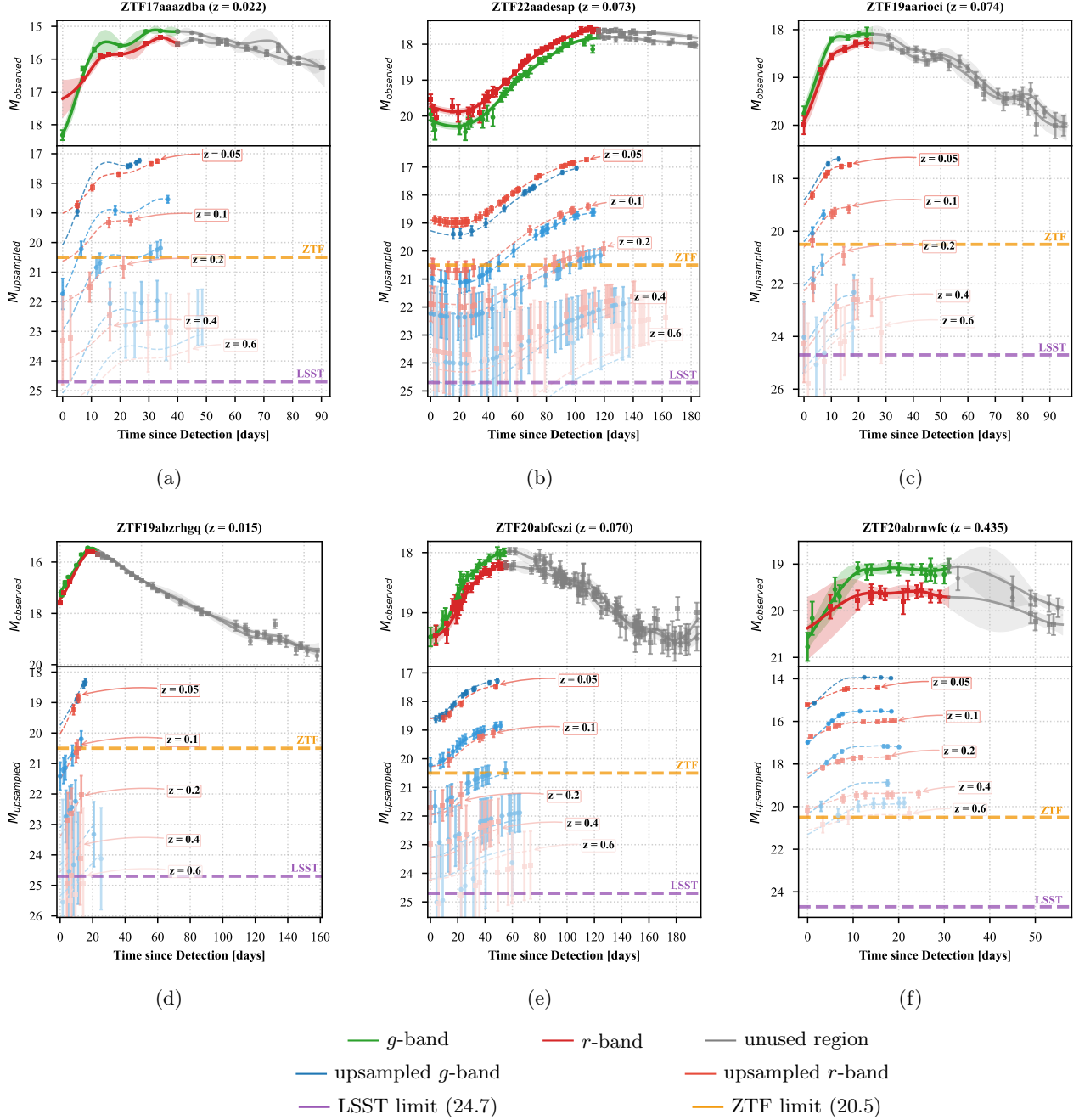
\begin{figure*}[htbp]
    \centering

    \begin{subfigure}[b]{0.3\textwidth}
        \includegraphics[width=\textwidth]{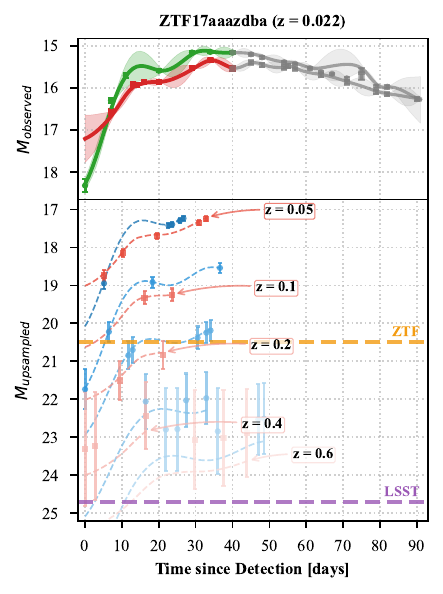}
        \caption{}
    \end{subfigure}
    \hspace{0.01\textwidth}
    \begin{subfigure}[b]{0.3\textwidth}
        \includegraphics[width=\textwidth]{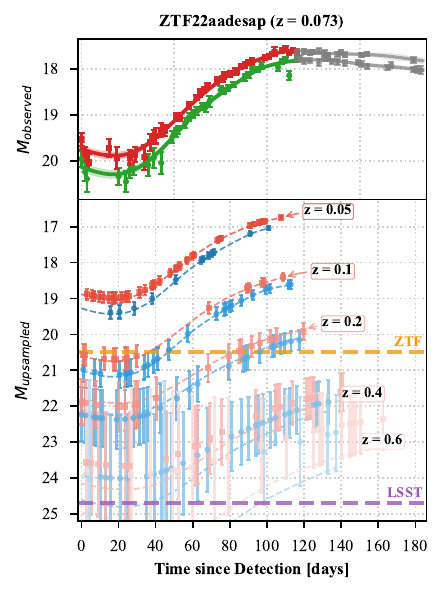}
        \caption{}
    \end{subfigure}
    \hspace{0.01\textwidth}
    \begin{subfigure}[b]{0.3\textwidth}
        \includegraphics[width=\textwidth]{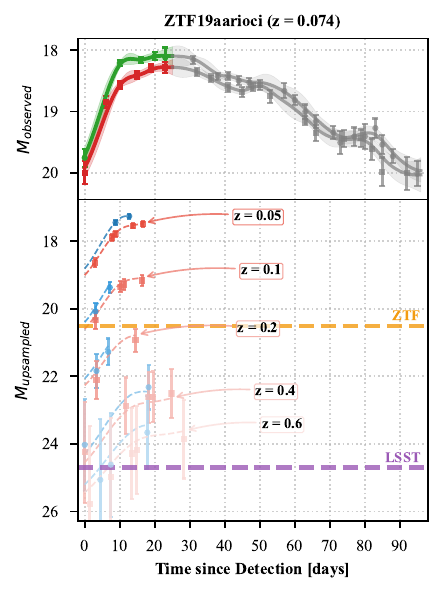}
        \caption{}
    \end{subfigure}

    \vspace{4pt} 

    \begin{subfigure}[b]{0.3\textwidth}
        \includegraphics[width=\textwidth]{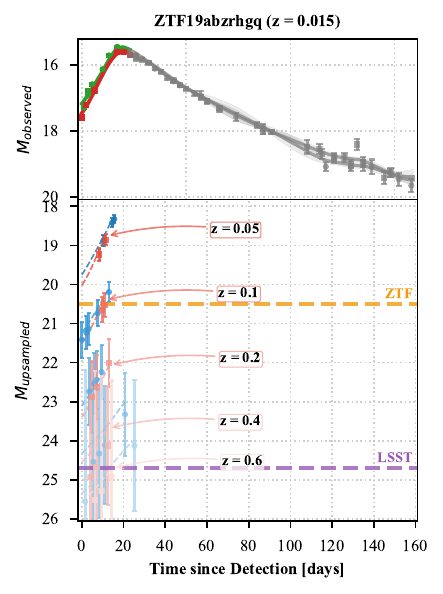}
        \caption{}
    \end{subfigure}
    \hspace{0.01\textwidth}
    \begin{subfigure}[b]{0.3\textwidth}
        \includegraphics[width=\textwidth]{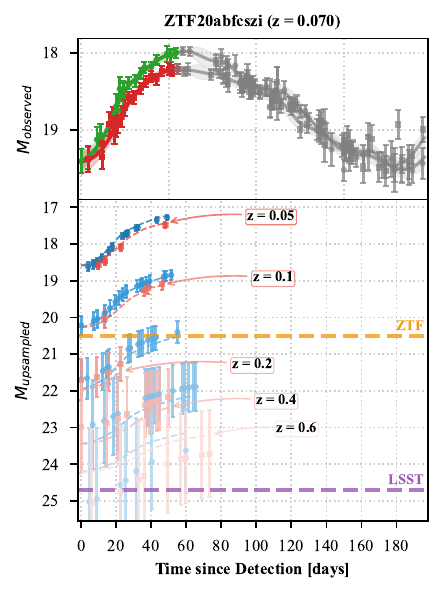}
        \caption{}
    \end{subfigure}
    \hspace{0.01\textwidth}
    \begin{subfigure}[b]{0.3\textwidth}
        \includegraphics[width=\textwidth]{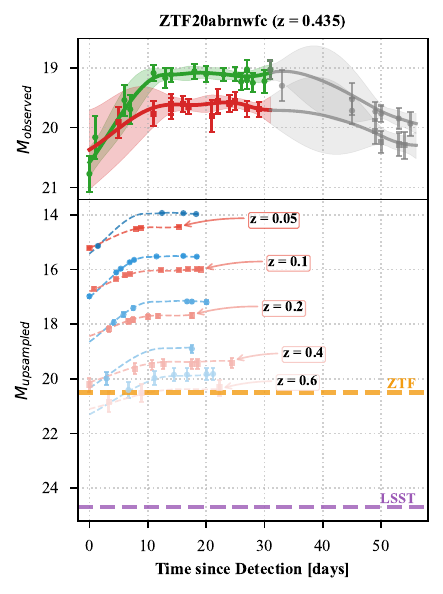}
        \caption{}
    \end{subfigure}
\vspace{10pt}
    \centering
    \begin{tabular}{@{}c@{\hspace{15pt}}c@{\hspace{15pt}}c@{\hspace{15pt}}c@{\hspace{15pt}}c@{}}
        \begin{tikzpicture}[baseline=-0.5ex]
            \draw[myGreen, thick] (0,0) -- (1.5em,0) node[right, black] {$g$-band};
        \end{tikzpicture} &
        \begin{tikzpicture}[baseline=-0.5ex]
            \draw[myMaroon, thick] (0,0) -- (1.5em,0) node[right, black] {$r$-band};
        \end{tikzpicture} &
        
        \begin{tikzpicture}[baseline=-0.5ex]
            \draw[gray, thick] (0,0) -- (1.5em,0) node[right, black] {unused region};
        \end{tikzpicture} \cr
        \begin{tikzpicture}[baseline=-0.5ex]
            \draw[myBlue, thick] (0,0) -- (1.5em,0) node[right, black] {upsampled $g$-band};
        \end{tikzpicture} & &
        \begin{tikzpicture}[baseline=-0.5ex]
            \draw[myRed, thick] (0,0) -- (1.5em,0) node[right, black] {upsampled $r$-band};
        \end{tikzpicture} \cr
        \begin{tikzpicture}[baseline=-0.5ex]
            \draw[myPurple, thick] (0,0) -- (1.5em,0) node[right, black] {LSST limit (24.7)};
        \end{tikzpicture} & &
        \begin{tikzpicture}[baseline=-0.5ex]
            \draw[myOrange, thick] (0,0) -- (1.5em,0) node[right, black] {ZTF limit (20.5)};
        \end{tikzpicture}
    \end{tabular}
    \caption{Six examples of resampled TDE light curves. Details are explained in Figure \ref{fig:slsn-lc-upsample}. }
    \label{fig:tde-lc-upsample}
\end{figure*}

The GP fitting applies to the full light curve but only the rising phase will be simulated for our training. Figures~\ref{fig:slsn-lc-upsample} and \ref{fig:tde-lc-upsample} depict this implementation for SLSN-I and TDE samples. This includes some examples where the GP fitting struggles in the fading phase due to insufficient detections, but the rising phase fitting is still useful.

\subsection{Early-phase  detection simulation}

Based on the 2D GP fitting, we simulate new detections by re-sampling the fitted light curves. As noted above, we restrict this process to the early phase, allowing sampling up to 10 days after the peak. This ensures that new detections can still include data points near the peak luminosity.
We exclude a few samples with poor GP fits (due to lack of detections in the rising phase, or over-/under-fitting to original data from the re-sampling and subsequent cross-matching procedures (section \ref{section5}), although their original observed data are still retained for training. 
The number of re-sampled detections during the rising phase is chosen randomly and can range from a single point up to the same number of original observations. We do not assume a uniform ZTF/LSST cadence here, since for individual objects at different coordinates the sampling is highly irregular.


\newpage

\subsection{Simulating transients at different redshifts}

When upsampling the rare objects in our training set, we wish to cross-match simulated light curves with host galaxy images at different redshifts (see section \ref{sec:crossmatching}).
Assuming a flat Universe with cold dark matter and a cosmological constant, with $\Omega_M=0.31$ and $H_0 = 67.7$\,km\,s$^{-1}$\,Mpc$^{-1}$ \citep{Planck2018_2020},
we derive the distance moduli at the redshift of the original light curve and at the redshift of the target galaxy, and use these to re-scale the apparent magnitude to the target redshift. This is implemented by \texttt{Astropy} cosmology tools. We adjust the photometric errors, given the new distance and survey sensitivity, using the method in Appendix \ref{sec:errors} (though note that errors are not currently used as a feature in \texttt{NEEDLE}). We also rescale the light curve width taking into account the difference in cosmological time-dilation at the original and new redshift.

\section{Cross-matched Samples}\label{sec:crossmatching}

\begin{figure*} 

\centering 
\includegraphics[width=\linewidth]{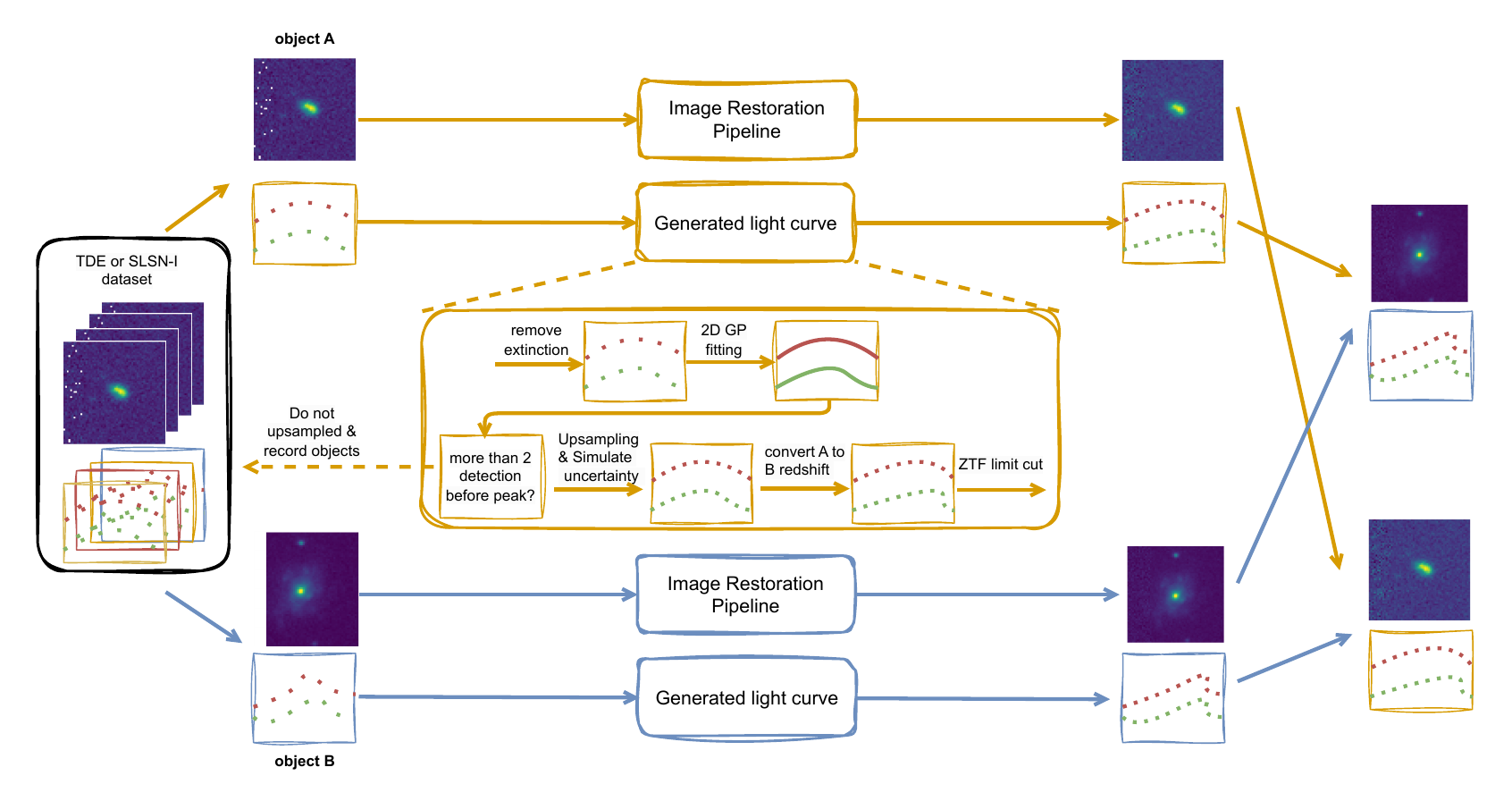} 
\caption{Light curve up-sampling pipeline. Here object A, B are two objects with the same label-SLSN-I or TDE, re-sampled and redshift-converted light curve from A is cross-mated with augmented images from B to form a new sample.} 
\label{fig:light_curve_pipeline} 

\end{figure*}

To augment training data for rare classes such as SLSNe-I and TDEs, we propose a cross-matching method that pairs host galaxy images with light curves of other objects from the same class, scaled to the galaxy redshift. This reduces the risk of over-fitting that would occur if we repeatedly generated the same light curve in the same galaxy. We assume that samples within each class are broadly representative of the underlying populations, and that there are no strong correlations between host and light curve properties \emph{within} a given class. By doing so, we can generate as many synthetic samples as needed to enhance model generalization and address class imbalance. The approach is shown in Figure~\ref{fig:light_curve_pipeline}.

This cross-matching strategy addresses both sample scarcity and intra-class diversity. 
To maintain astrophysical consistency, redshifts are sampled to follow the observed redshift distribution of the corresponding class. For each synthetic sample, we randomly select an existing light curve and pair it with a host galaxy from another object within the same class. The redshift range for cross-matching is restricted to within $\pm20\%$ of the original redshift, ensuring the synthetic sample distribution remains similar to that of the real dataset, and avoids the need to estimate K-corrections \citep{Hogg_2002}.

Moreover, this ensures that cross-matching hosts across redshifts does not significantly distort population characteristics. For example, while SLSN-I hosts are typically faint, low-mass and low-metallicity dwarf galaxies, they can also appear in brighter systems at high redshifts, where younger galaxies are less metal-rich for the same mass. For both SLSNe and TDEs, the ZTF sample are typically at $z\ll1$, so that no strong redshift evolution of host galaxy properties is expected within this sample. Restricting to cross-matches within a narrow redshift slice will ensure that the approach still works for deeper surveys (such as LSST), where cosmological evolution in host properties may be observable.

This approach also mitigates the effects of Malmquist bias in the training set, where events detected at higher redshift are more likely to come from the bright tail of the magnitude distribution. Cross-matching these events with low-redshift galaxies would lead to an over-abundance of very luminous events in our simulated data.
Converted light curves are constrained by the ZTF detection limit (20.5 mag); any re-sampled detections fainter than this threshold are discarded. 

\begin{figure*}[t]
    \centering
    \begin{subfigure}[t]{0.48\linewidth}
        \centering
        \includegraphics[width=\linewidth]{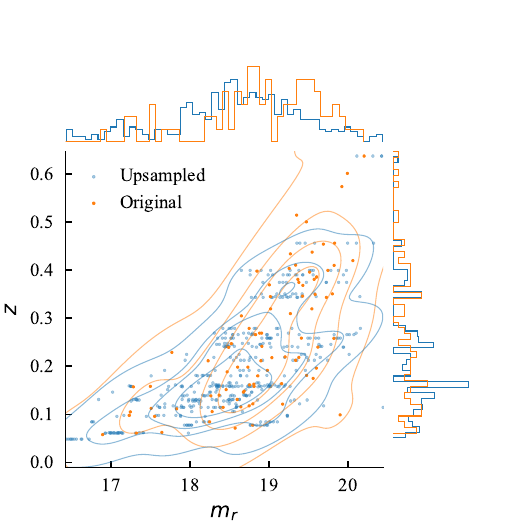}
        \caption{Scatter plot of redshift and apparent magnitude in $r$ band for SLSN-I class in original (orange) and upsampled (blue) sets. }
        \label{fig:cross_slsn}
    \end{subfigure}
    \hfill
    \begin{subfigure}[t]{0.48\linewidth}
        \centering
        \includegraphics[width=\linewidth]{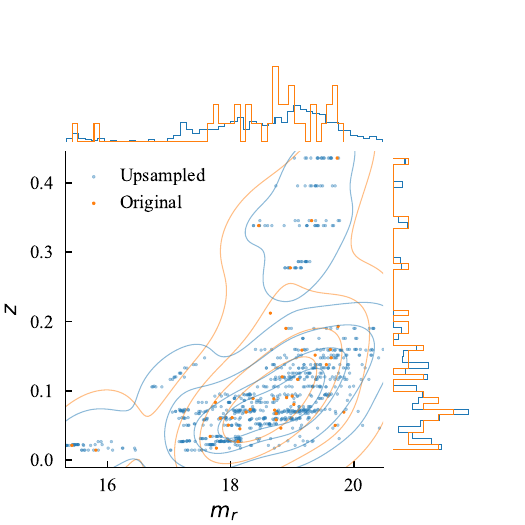}
        \caption{Scatter plot of redshift and apparent magnitude in $r$ band for TDE class in original (orange) and ups-ampled (blue) sets.}
        \label{fig:cross_tde}
    \end{subfigure}
    \caption{Comparison of redshift–apparent magnitude distributions for (a) SLSN-I and (b) TDE classes in original vs. up-sampled sets. The initial up-sample number for SLSNe-I and TDEs is 2000 for each, the number is reduced after ZTF limit cut-off and other selection criteria.}
    \label{fig:cross_slsn_tde}
\end{figure*}

Finally, we acknowledge that observational biases can introduce artificial correlations between transient properties and their host galaxies. This issue has been discussed in previous studies (e.g., \citealt{Lunnan_2014}), and our cross-matching and augmentation method is designed with an awareness of these potential effects.
The up-sampled SLSN-I and TDE populations are illustrated in Figure \ref{fig:cross_slsn_tde}. In both plots, we observe that as the redshift increases and the apparent magnitude becomes fainter, the discovery rate declines, which is an expected result of survey observing limits. We can see the same effect is captured in our synthetic data, with the bulk of detections at low redshift.

\section{Training and model performance}\label{section5}

With the data augmentation complete, we now retrain our classifier. In the original version of \texttt{NEEDLE}, we prioritised completeness in recovering rare classes. As we transition towards LSST and the higher rates of transient discovery, we now focus on purity, to minimise the demand on spectroscopic follow-up resources.

\subsection{Customized focal loss}

To further enhance the sample purity, we incorporated a focal loss (FL) term into our original class-weighted cross-entropy (CE) objective \citep{Sheng_2024}. The focal loss mitigates class imbalance by down-weighting well-classified (high-confidence) examples and relatively up-weighting hard (low-confidence) examples during training \citep{Lin_2017}. 

This modification discourages over-confident predictions and substantially improves the purity of the predicted classes in imbalanced datasets, albeit at the cost of reduced completeness relative to the standard CE loss. The behavior of the combined FL+CE loss is also influenced by class-dependent feature-learning dynamics: in our testing, we found that when applying the same FL+CE hyperparameter configuration across all labels, the completeness for TDEs saturates (reaches a plateau) after fewer training epochs than that for SLSNe-I.
To achieve an appropriate balance between sample purity and class completeness for SLSNe-I and TDEs, we refine the FL by assigning different parameters for each class. The resulting loss function is 
\begin{align}
FL(p_t) &= \sum -\alpha_{i} (1 - p_{t,i})^{\gamma_{i}} \log(p_{{t,i}}), 
\label{eq:focal-loss}
\end{align}

where $i$ is the label index. 0: SN, 1: SLSN-I, 2: TDE; $t$ refers to one sample with label $i$; $p_{t,i}$ is the predicted probability of sample $t$ for the true class $i$.; $\alpha_i$ is the class-specific weighting factor to balance positive and negative classes.; $\gamma_{i}$ is the focusing parameter for the true class $i$, which controls the down-weighting of easy examples.
The optimized parameter values for each class are summarized in Table~\ref{tab:focal_params}.

\begin{table}
\centering
\resizebox{.25\columnwidth}{!}{%
\begin{tabular}{lll}
\hline
\multicolumn{1}{c}{\textbf{label}} & \multicolumn{1}{c}{\textbf{$\gamma$}} & \multicolumn{1}{c}{\textbf{$\alpha$}} \\ \hline
SN     & 0.0   & 0.02 \\ \hline
SLSN-I & 2.0 &  0.49 \\ \hline
TDE    & 2.5   &  0.49  \\ \hline

\end{tabular}%
}
\caption{The optimal focal loss parameter set.}
\label{tab:focal_params}
\end{table}

\subsection{Calibration methods}

We employ the stratified sampling method \citep{Merrillees_2021} to split training and validation sets, to ensure that they have the same class distribution as the original dataset.
Random samples are drawn from each subgroup, and 20\% of the samples from each label set are drawn in proportion to its size in the population to form a validation set.
We up-sample the training set using the cross-matching augmentation procedure described in Section~\ref{sec:crossmatching}, but do not include any synthetic samples in the validation set. 

We employ a K-fold cross-validation strategy \citep{Kohavi_1995}, in which the combined training and validation sets are randomly partitioned into 10 folds (to train 10 models individually), and performance is averaged over all folds. 
The model with performance closest to the average will be selected for real-time deployment in \texttt{NEEDLE}.

Across all datasets, objects with no detected host galaxy are stored separately and used to train a separate model for hostless transients -- we refer to this as \texttt{NEEDLE-T} (our default model with transient and host information is \texttt{NEEDLE-TH}). In this work, we report results only for hosted objects, as the cross-matching procedure is primarily advantageous for informative images containing host galaxies, and this subset includes both SLSN-I and TDE classes. We emphasise that we can still classify hostless objects as in \citet{Sheng_2024}, and will continue to refine the model for these also.

To preserve the model’s generalization capability, the up-sampling ratio should not be excessively large; otherwise, the model may overfit to noise in the training set. We therefore apply 200 up-samples for SLSNe-I and TDEs. Experiments show that this setting yields significant improvements in both recall and precision compared to the model without up-sampling. 

\subsection{Untouched test set}

To verify that the model performs well on unseen events, and not just those in the training and validation sets, we include an additional test set of untouched ZTF transients following \citet{Sheng_2024}. Compared with our previous work, the untouched set has been updated to include new objects up to June 2025. All samples were selected from the BTS survey, explicitly excluding any sources that were part of the training set (including both the initial dataset and the untouched set described in \citealt{Sheng_2024}). This yielded a total of approximately 3,000 sources. After applying standard input selection criteria -- requiring rising detections and at least one pair of detection and reference images -- the final sample consists of 2,588 sources. The distribution of labels in this updated untouched set is summarized in Table~\ref{tab:untouched-set}. This updated dataset is used to estimate the completeness and purity of SLSN-I and TDE predictions that we could achieve with ZTF real alerts.

\begin{table}
\centering
\resizebox{.4\columnwidth}{!}{%
\begin{tabular}{llll}
\hline
\multicolumn{1}{c}{\textbf{label}} & \multicolumn{1}{c}{\textbf{hosted}} & \multicolumn{1}{c}{\textbf{hostless}} & \textbf{sum} \\ \hline
SLSN-I & 10    & 4   & 14   \\ \hline
TDE    & 15    & 0   & 15   \\ \hline
SN     & 2391 & 168 & 2559 \\ \hline
sum    & 2416  & 172 & 2588 \\ \hline
\end{tabular}%
}
\caption{Updated untouched set by June 2025.}
\label{tab:untouched-set}
\end{table}

\subsection{Ablation Experiments}\label{sec:ablation}

Here, we present the ablation analysis demonstrating the incremental gains provided by masking and up-sampling. The evaluation metrics we use to compare models are defined in Table \ref{tab:terminology}, with a focus on enhancing the discovery of SLSNe-I and TDEs while reducing SN contamination.

\begin{table}[]
\centering
\begin{tabular}{c p{12cm}}
\hline
\textbf{Metric}       & \multicolumn{1}{c}{\textbf{Explanation}}                                                                                                                                                                                                                               \\ \hline
macro-confidence  & The unweighted average predicted probability of the preferred label across all three classes.                                                                                           \\
b-confidence   &   Biased confidence. The unweighted average predicted probability the preferred label for only the SLSN-I and TDE classes.                                                                                                                                                                         \\
b-entropy   & Biased entropy. The unweighted average difference between the predicted probabilities and the true labels for SLSN-I and TDE.                                                                                                                                                                            \\

Completeness (Recall) & Proportion of actual positives correctly identified: $\text{TP} / (\text{TP} + \text{FN})$.  
\\
Purity (Precision)    & Proportion of predicted positives that are true positives: $\text{TP} / (\text{TP} + \text{FP})$. 
    \\
F1 Score              & Harmonic mean of precision and recall; $F1 = 2 \times \frac{\text{Precision} \times \text{Recall}}{\text{Precision} + \text{Recall}}$. \\

Coverage & The number of samples in a class that remain for decisive predictions after discarding those whose highest predicted probability falls below the confidence threshold.
\\

\hline
\end{tabular}
\caption{Terminology explanation for metrics presented in Section \ref{sec:ablation}. TP = True Positive, FP = False Positive, TN = True Negative, FN = False Negative.}
\label{tab:terminology}
\end{table}

Table \ref{tab:metrics} reports the averaged scores for both the validation and untouched sets across 10 K-fold models. We first look at the model confidence, i.e.~how high a probability it assigns to the preferred class. Although the overall confidence changes only marginally across model variants, the \emph{class-specific} b-confidence for SLSNe-I and TDEs increases substantially with the inclusion of masking and up-sampling. 
This means that the augmented model is more likely to assign higher probability scores to examples of these rarer classes, compared to the scores assigned by the default model. 
Correspondingly, the b-entropy scores decrease in a consistent manner for both datasets. These results indicate that masking and up-sampling enhance the model’s ability to assign higher confidence to rare events, which is an effect that is particularly valuable if we want to set a confidence threshold in our classifications to more effectively remove false positives.

\begin{table}[t]
    \centering
    \begin{subtable}[t]{0.8\textwidth}
        \centering   
        \begin{tabular}{ccccc}
\hline
masking               & upsampled & macro-confidence & b-confidence   & b-entropy      \\ \hline
False                 & False     & 0.868 +/-0.016   & 0.64 +/-0.012  & 0.727 +/-0.017 \\ \cline{1-5} 
\multirow{2}{*}{True} & False     & 0.878 +/-0.007   & 0.678 +/-0.022 & 0.665 +/-0.019 \\ \cline{2-5} 
                      & True      & 0.873 +/-0.017   & 0.734 +/-0.032 & 0.564 +/-0.039 \\ \hline
\end{tabular}
\caption{Validation set metrics}
    \end{subtable}
    \hfill
    \begin{subtable}[t]{0.8\textwidth}
        \centering
        \begin{tabular}{ccccc}
\hline
masking               & upsampled & macro-confidence & b-confidence   & b-entropy      \\ \hline
False                 & False     & 0.874 +/-0.017   & 0.638 +/-0.024 & 0.744 +/-0.03  \\ \cline{1-5} 
\multirow{2}{*}{True} & False     & 0.88 +/-0.007    & 0.674 +/-0.022 & 0.685 +/-0.036 \\ \cline{2-5} 
                      & True      & 0.878 +/-0.015   & 0.746 +/-0.019 & 0.545 +/-0.023 \\ \hline
\end{tabular}
  \caption{Untouched set metrics}
    \end{subtable}
\caption{Metrics for the validation (a) and untouched sets (b) for each model. Each cell reports the mean and standard deviation of the score across 10 K-fold models. }
\label{tab:metrics}
\end{table}

\begin{figure}
    \centering
    \includegraphics[width=\linewidth]{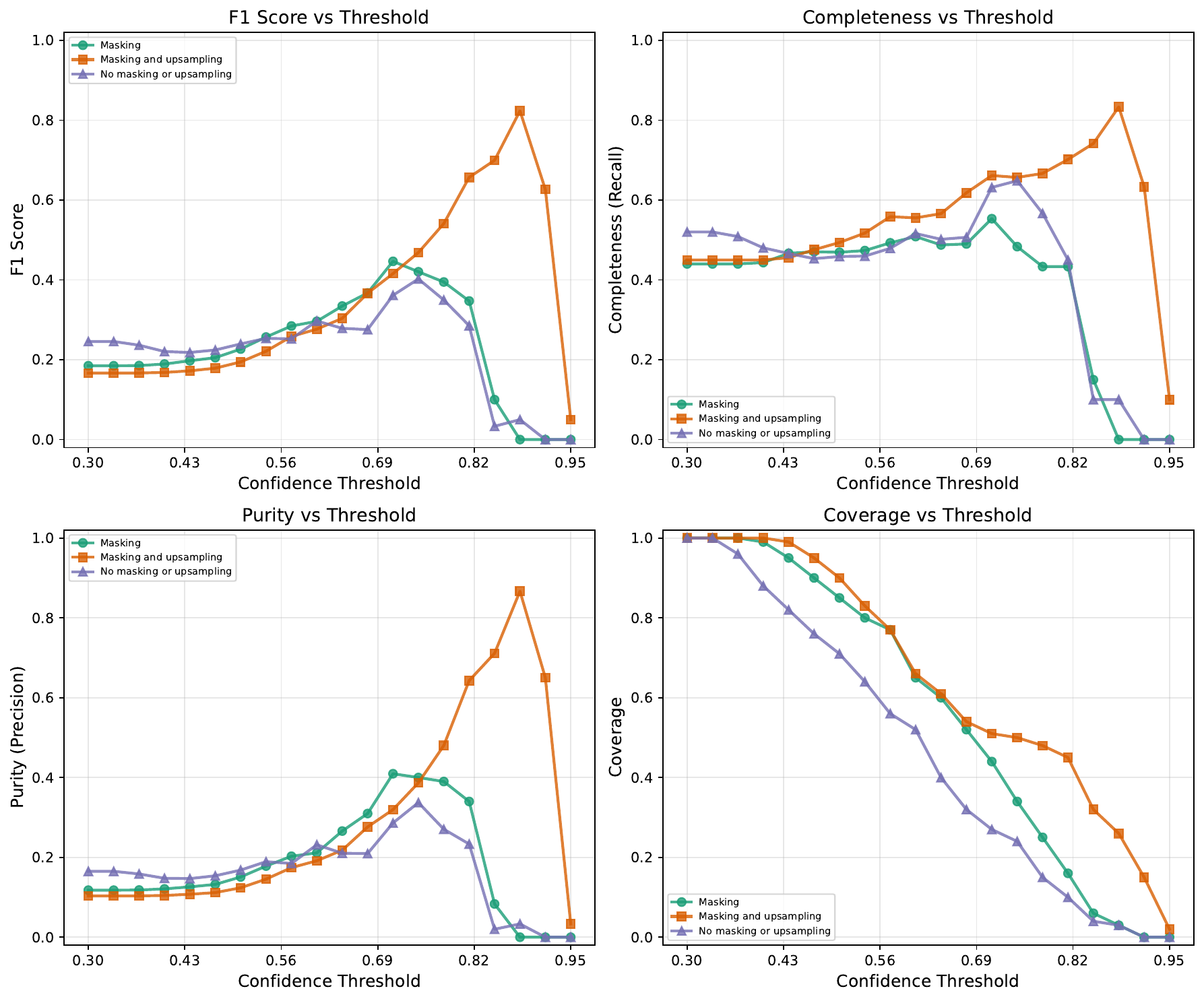}
    \caption{F1, completeness, purity, and coverage metrics for SLSNe-I in the untouched set, evaluated using the \texttt{NEEDLE-TH} model trained with data augmented under the `Masking', `Masking and upsampling', and `No masking or upsampling' configurations. Metric definitions are given in Table~\ref{tab:terminology}.
}
    \label{fig:untouched-slsn-scores}
\end{figure}

\begin{figure}
    \centering
    \includegraphics[width=\linewidth]{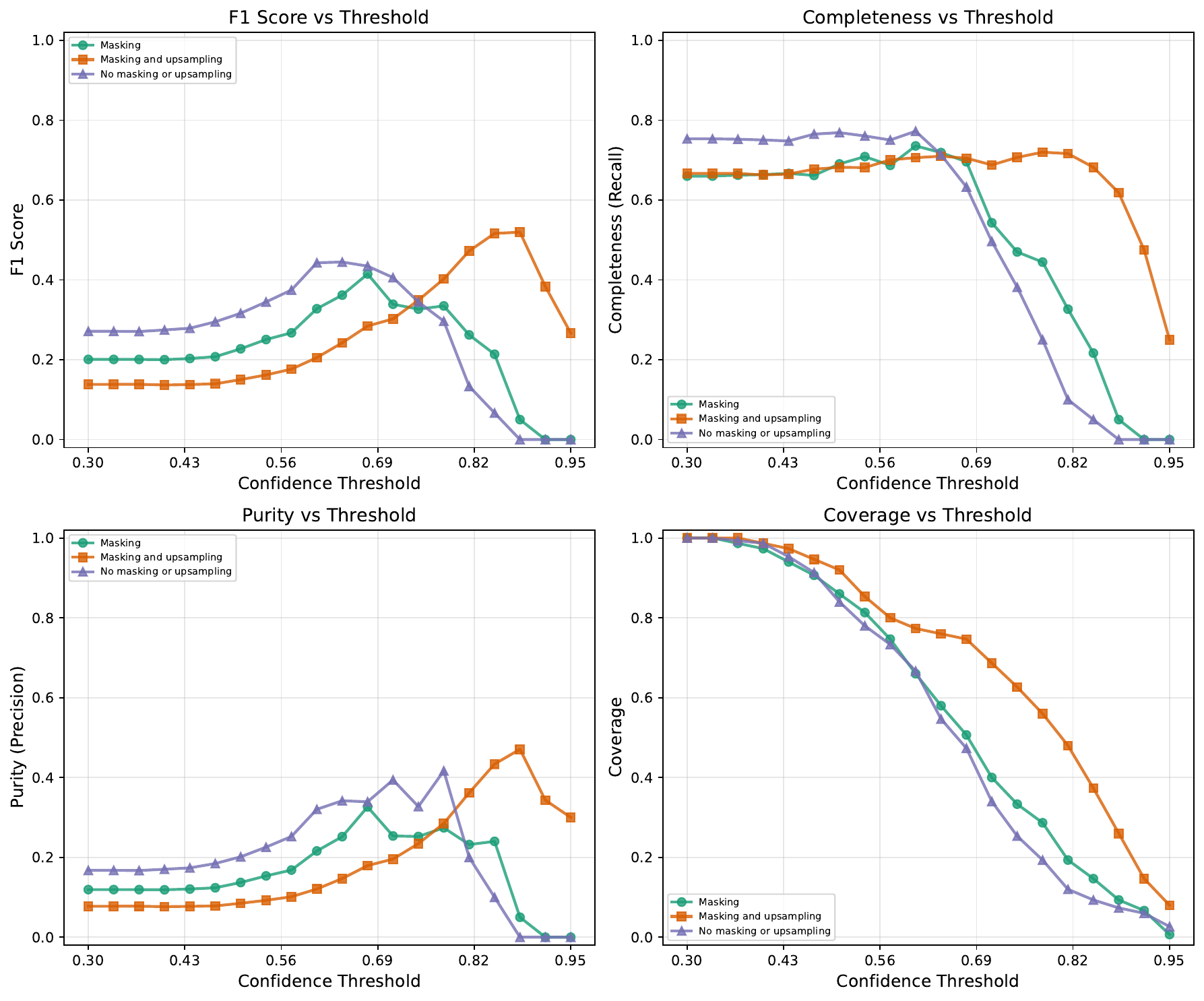}
    \caption{F1, completeness, purity, and coverage metrics for TDEs in the untouched set.}
    \label{fig:untouched-tde-scores}
\end{figure}

Figures \ref{fig:untouched-slsn-scores} and \ref{fig:untouched-tde-scores} show the trends of F1 score, completeness, purity, and coverage as functions of the confidence threshold, averaged over 10 K-fold models for SLSN-I and TDE samples in the untouched set, respectively. The scores are computed using the untouched samples whose mean maximum probabilities exceed the specified threshold.
All models are implemented using the FL loss function.

For SLSN-I samples, the Masking configuration outperforms the original setup (without masking or up-sampling) in terms of F1 score, purity, and coverage, suggesting that masking allows the model to more effectively focus on the faint host galaxies. This improvement likely accounts for the observed decrease in completeness within the confidence threshold range of $0.7\sim0.8$. Specifically, normal SNe in faint host galaxies share similar contextual characteristics to SLSNe-I, which leads to some SLSNe-I being misclassified as SNe and receiving lower scores.

The orange curves, corresponding to the configuration with both masking and up-sampling, always achieves higher coverage, but can exhibit lower purity and completeness than the other configurations at low confidence thresholds. However, the masked and upsampled model surpasses the others once the threshold exceeds $\approx0.8$. 
This indicates that the model achieves relatively higher confidence on true positives. This effect is especially strong for TDEs. The high purity achieved above a threshold of 0.8, especially for SLSNe-I, significantly reduces the amount of contaminants and the time needed for expert eyeballing. 

\begin{figure}
    \centering
    \includegraphics[width=\linewidth]{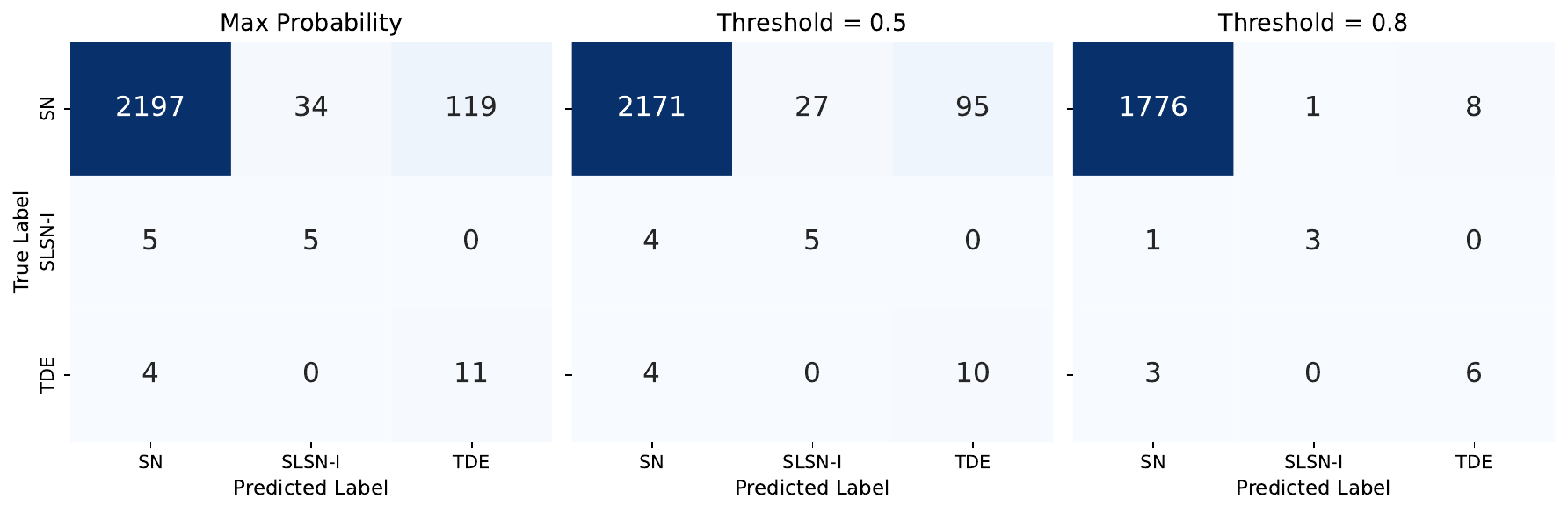}
    \caption{Confusion matrix at threshold maximal, 0.5 and 0.8 for NEEDLE-TH models with configuration of masking, upsampling and focal loss.}
    \label{fig:best-model-untouched}
\end{figure}

The confusion matrices for the masked and upsampled configuration are shown in Figure \ref{fig:best-model-untouched}. At a threshold of 0.8, the coverage drops to 4 out of 10 SLSNe-I and 9 out of 15 TDEs. Although this is not ideal in terms of overall completeness, the resulting purities, 0.75 (3/4) for SLSNe-I and 0.43 (6/14) for TDEs, meaningfully improve the efficiency and success rate of selecting follow-up targets. This is particularly valuable when dealing with millions of Rubin alerts per night. For comparison, the original version of \texttt{NEEDLE} achieved a purity of $\sim 0.2$ for both classes \citep{Sheng_2024}. 
In Appendix \ref{sec:fl}, we compare our best-performing models with counterparts trained without focal loss. We find that the modified loss function plays a crucial role in determining model purity, with focal-loss–based models significantly outperforming those trained without FL across confidence thresholds in the range 0.7–0.9. By contrast, models without FL provide more favorable configurations for achieving higher completeness and coverage.

\section{Conclusions}\label{section6}

This paper introduces a novel data augmentation pipeline for images and light curves, designed to address the extreme class imbalance problem in identifying rare transients in time-domain surveys. 

To mitigate the impact of image artefacts in the \textit{Science} and \textit{Reference} frames, we employ a Similarity Index to identify abnormal pixels and replace them with appropriately scaled pixels from other images at the same sky coordinates. Furthermore, because the images are often dominated by bright background sources, Convolutional Neural Networks (CNNs) tend to overlook the faint, blue hosts of SLSNe-I. To address this, we implement a masking procedure that removes most irrelevant background objects while retaining the transient and its potential host within the frame. This approach also facilitates flexible image augmentation and upsampling: provided the host is fully contained within a $60\times60$ pixel frame, the image can be randomly rotated by arbitrary angles, preventing the model from learning spurious orientation-dependent features. Notably, the method is based entirely on real images, without simulations of transient-host morphology or color, thereby avoiding the introduction of artificial biases.

For light curves, we fit well-detected rising phases using a two-dimensional Gaussian Process, which captures correlations across both time and wavelength dimensions. By resampling detections, shifting to different redshifts, and recalculating measurement uncertainties, we generate a number of high-quality synthetic light curves for SLSNe-I and TDEs. A key innovation of this pipeline is a new upsampling strategy: within a single label set, we cross-match the processed images of one object with the upsampled, redshift-adjusted light curve of another to produce new, realistic samples. This approach yields redshift and magnitude distributions that closely resemble those of the original data.

In addition, we introduce a modified focal loss function specifically designed to enhance the purity of the SLSNe-I and TDE classes while preserving overall classification confidence. This approach is broadly applicable to classification tasks targeting other rare transient populations in highly imbalanced datasets.

Our models configured with masking and up-sampling show a large increase in purity above a confidence threshold of 0.8 for both SLSNe-I and TDE samples in an unseen data set. Although the coverage of the true positives decreases, the resulting high purity values, 0.75 for SLSNe-I and 0.43 for TDEs, will significantly speed up the follow-up process by achieving higher success rates. In the production version \texttt{Lasair-NEEDLE}, running on real-time alerts, we will also provide predictions from models that prioritize high completeness, as in our earlier work \citep{Sheng_2024}. 
While the high-completeness model produces a manageable number of candidates in ZTF, at the data rates of LSST the model that prioritizes purity may be more appropriate.

Overall, experimental results demonstrate that the proposed pipeline substantially enhances both the robustness and representativeness of the training dataset. In particular, it markedly improves the purity of rare transient classes in real-time alert systems, offering a more reliable basis for early classification.

\section{Acknowledgments}
MN and XS are supported by the European Research Council (ERC) under the European Union’s Horizon 2020 research and innovation programme (grant agreement No.~948381).

\newpage
\bibliography{main}{}
\bibliographystyle{aasjournalv7}

\appendix

\section{Method to simulate magnitude errors} 
\label{sec:errors}
After redshift conversion, we simulate the magnitude error following below formulas: 

\begin{align}
    & \text{\textbf{(1) Poisson error on flux:}} \nonumber \\[0.5ex]
    & \quad F_{\text{err}} = \sqrt{F} \quad \\
    & \text{\textbf{(2) Magnitude error approximated as the fractional flux error:}} \nonumber \\[0.5ex]
    & \quad m_{\text{err}} \approx \frac{F_{\text{err}}}{F} = \frac{\sqrt{F}}{F} = \frac{1}{\sqrt{F}} \quad  \\
    & \text{\textbf{(3) Conversion of magnitude to flux to express the equation in terms of magnitude:}} \nonumber \\[0.5ex]
    & \quad F = \text{constant} \cdot 10^{-0.4m} \quad  \\
    & \text{\textbf{(4) Substituting formula in (3) into (2):}} \nonumber \\[0.5ex]
    & \quad m_{\text{err}} \approx \frac{1}{\sqrt{\text{constant} \cdot 10^{-0.4m}}} = k \cdot 10^{0.2m}, \quad \text{where $k$ is a constant} \quad \\
    & \text{\textbf{(5) Determining the constant using the ZTF detection limit}} \nonumber \\[0.5ex]
    & \quad \text{(3$\sigma$ corresponds to a magnitude error of $\sim0.3$):} \nonumber \\[0.5ex]
    & \quad \text{When } m = 20.5, \, m_{\text{err}} = 0.3 \\
    & \text{\textbf{(6) Substituting $m = 20.5$ into (4) and solving for $k$:}} \nonumber \\[0.5ex]
    & \quad 0.3 = k \cdot 10^{0.2 \cdot 20.5} \implies k = \frac{0.3}{10^{4.1}} \approx 2.4 \times 10^{-5} \\
    & \quad \textbf{Thus, the final relation is:} \nonumber \\[0.5ex]
    & \quad m_{\text{err}} = 2.4 \times 10^{-5} \cdot 10^{0.2m} \\
    & \textbf{(7) Here is how we apply $m_{\text{err}}$ to magnitude to simulate the uncertainty of observation:} \nonumber \\[0.5ex]
    & \quad \text{Perturbing re-sampled magnitudes $m_{\text{err,i}}$ with a random factor $\epsilon_i$:} \nonumber \\[0.5ex]
    &\quad m_i' = m_i + m_{\text{err,i}} \cdot \epsilon_i,\quad \epsilon_i \sim \mathcal{U}(-1,\ 1)  \\
    & \quad \text{Resulting in the final re-sampled light curve with simulated uncertainty:} \nonumber \\[0.5ex]
    &\quad \mathbf{m'} = \left[ m_1 + \sigma_1' \cdot \epsilon_1,\ m_2 + \sigma_2' \cdot \epsilon_2,\ \ldots \right] \quad  \\ \nonumber
\end{align}

\section{The importance of the focal loss}\label{sec:fl}

\begin{figure}
    \centering
    \includegraphics[width=\linewidth]{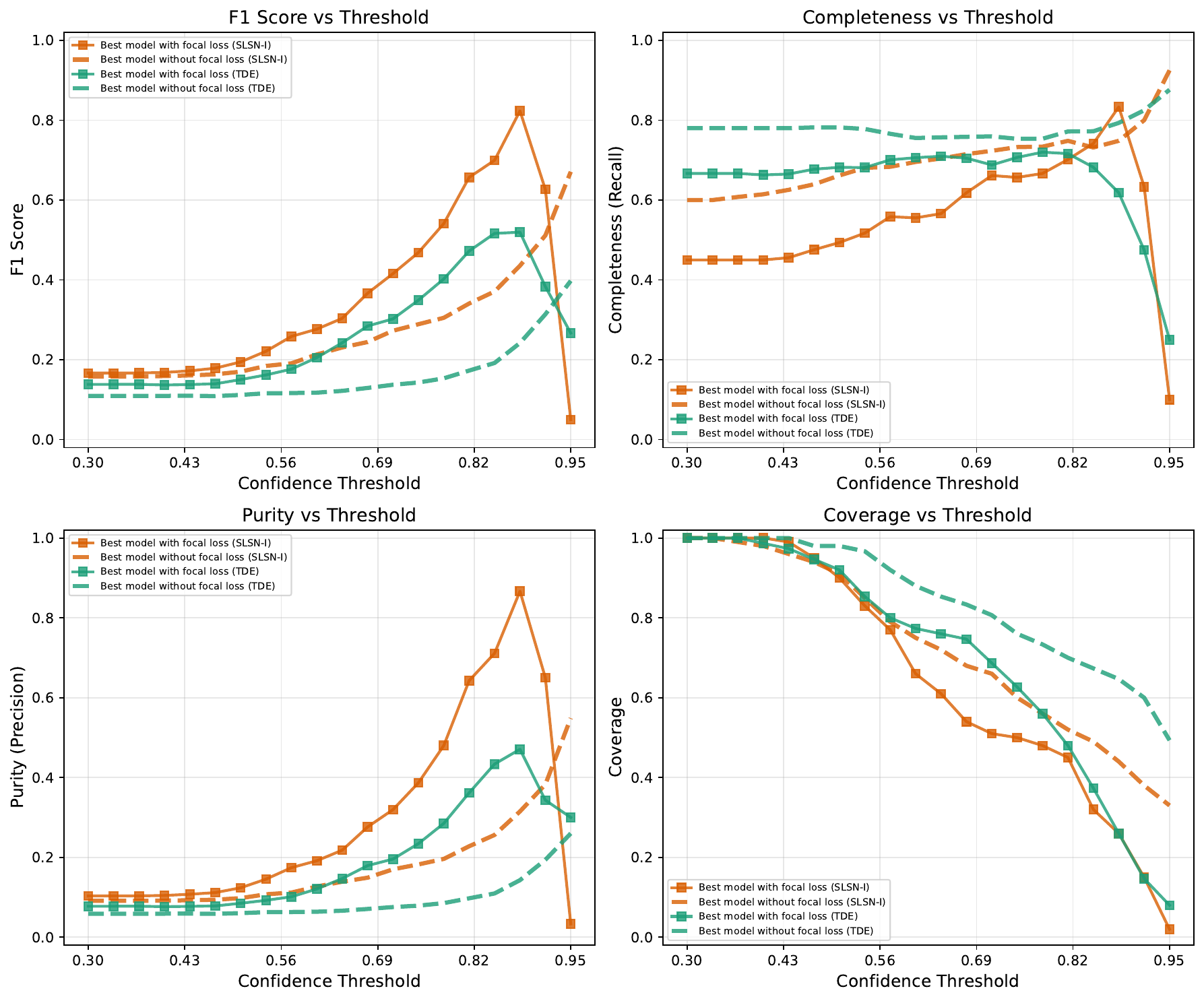}
    \caption{The metrics on the untouched set are obtained from models trained with masking and up-sampling, both with and without the focal loss. All values are averaged over 10 K-fold models.}
    \label{fig:app-focal-plots}
\end{figure}

Figure \ref{fig:app-focal-plots} presents a comparison of four metrics between our optimal models with modified focal loss and versions trained without focal loss. The results indicate that incorporating focal loss has a substantial impact on model purity, yielding markedly superior performance over confidence thresholds of 0.7$\sim$0.9. In contrast, models trained without FL are more effective when prioritizing higher completeness and coverage. 
In practice, prioritizing high-purity models for Rubin alerts will significantly shorten the candidate list and reduce the eyeballing time.

\end{document}